\documentclass[twocolumn, prl, reprint, amsfonts, amsmath, amssymb, superscriptaddress, longbibliography, floatfix,aps]{revtex4-1}
\usepackage[T1]{fontenc}
\usepackage{graphicx}
\usepackage[dvipsnames]{xcolor}
\usepackage[colorlinks,linkcolor=Blue,citecolor=Blue]{hyperref}
\usepackage{braket}
\usepackage{enumitem}
\usepackage[english]{babel}
\usepackage{titlesec} 

\begin{document}
\title{Superconducting Berry Curvature Dipole}
\author{Oles Matsyshyn}
\email{oles.matsyshyn@ntu.edu.sg}
\affiliation{Division of Physics and Applied Physics, School of Physical and Mathematical Sciences, Nanyang Technological University, Singapore 637371}
\author{Giovanni Vignale}
\affiliation{The Institute for Functional Intelligent Materials (I-FIM),
National University of Singapore, 4 Science Drive 2, Singapore 117544}
\author{Justin C. W. Song}
\email{justinsong@ntu.edu.sg}
\affiliation{Division of Physics and Applied Physics, School of Physical and Mathematical Sciences, Nanyang Technological University, Singapore 637371}
\date{\today}
 
\begin{abstract}
Superconductivity and Bloch band Berry curvature responses represent two distinct paradigms of quantum coherent phenomena. The former relies on the collective motion of Cooper pairs while the latter proceeds from the momentum-space winding of Bloch wave functions. Here we reveal a superconducting Berry curvature dipole (BCD) that arises as a collective phenomenon in noncentrosymmetric superconductors. Strikingly, we find the superconducting BCD is sensitive to the {\it phase} of the order parameter and depends on the noncentrosymmetric structure of its pairing. This unusual property enables a BCD proximity effect in hybrid quantum materials that induces nonreciprocity even in a target centrosymmetric metal. We find a superconducting BCD naturally produces nonreciprocal electromagnetic responses that include dissipationless supercurrent-induced dynamical Hall conductivity as well as a giant second-order nonlinearity. This renders noncentrosymmetric superconductors an exciting platform for realizing unconventional dissipationless responses and their BCD responses a novel diagnostic of the structure of the superconducting gap. 
\end{abstract}

\maketitle

Bloch band geometry plays a critical role in realizing novel types of electronic dynamics in quantum matter \cite{RevModPhys.82.1959}. A prime example is the Berry curvature dipole (BCD)~\cite{SodemannFu} that endows electrons with a range of nonreciprocal responses that include nonlinear Hall currents even in the presence of time-reversal symmetry \cite{SodemannFu,ma2019observation,Kang2019}, electro-optic modulation~\cite{PhysRevB.99.155404}, and amplification without population inversion \cite{PhysRevB.107.125151,PhysRevLett.130.076901}. Much like other quantum geometrical quantities \cite{Ahn2022}, the BCD is often viewed as an intrinsic characteristic of individual Bloch electrons \cite{matsyshyn2019nonlinear}. Even in superconductors, BCD can emerge from the quantum geometry of Bogoliubov quasiparticles~\cite{Liang_2017,PhysRevLett.126.187001} yielding superconducting nonlinear optical responses~\cite{Yanase1,Yanase2}.

Here, we argue that the BCD in superconductors can take on an {\it emergent} character distinct from that of normal metallic states \cite{SodemannFu, matsyshyn2019nonlinear}. In particular, we find that a crucial contribution to the superconducting BCD arises from geometric phases accrued by the order parameter as the supercurrent flows. We term such collective BCD ``{\it order parameter BCD}''. Critically, order parameter BCD directly depends on the internal structure of Cooper pairing producing a total superconducting BCD that can be sharply different from that of its metallic parent state.  This leads us to intriguing predictions: e.g., BCD can be proximitized in a centrosymmetric metal (where normal metallic BCD vanishes) by a noncentrosymmetric superconducting substrate. Even in noncentrosymmetric parent materials with Bloch electrons that possess an intrinsic Bloch BCD, we find that order parameter BCD can be pronounced, rendering it essential in describing superconducting quantum geometry.

At a fundamental level, we trace this emergent behavior to a novel Cooper pair shift. Much as two electrons experience a collective displacement when they scatter off each other \cite{ThePesin, RoniIllan}, the Cooper pair shift tracks a collective displacement of the center-of-mass of two electrons as they Cooper-pair. In time-reversal symmetric superconductors, we find the Cooper pair shift closely tracks the inversion breaking nature of the order parameter, vanishing for centrosymmetric pairing. As we argue below, the Cooper pair shift describes an emergent length scale in noncentrosymmetric superconductors that determines the quantum geometry of superconductors such as the order parameter BCD. This produces an emergent Hall inertia when electrons Cooper pair distinct from that found for noninteracting electrons~\cite{matsyshyn2019nonlinear}. 

We expect pronounced order parameter BCD arises and dominates the BCD in a range of noncentrosymmetric superconductors, e.g., rhombohedral trilayer graphene (RTG)~\cite{Zhou2021}. In these, superconducting BCD can be directly probed through a range of nonreciprocal electromagnetic responses below the superconducting gap, e.g., a {\it nondissipative} supercurrent-induced dynamical Hall conductivity as well as superconducting BCD second-order nonlinearity. These are particularly exciting because such responses in metals are often dissipative~\cite{PhysRevB.99.155404,matsyshyn2019nonlinear,SodemannFu}.

\textit{Superconducting Berry curvature dipole.} We begin by examining an electronic system subject to an applied electromagnetic field, $\mathbf{A}(t)$; here $\mathbf{A}(t)$ is the vector potential. This can be tracked via the set of electronic states $\ket{\Phi(\mathbf{A})}$ and their energy $\mathcal{E}(\mathbf{A})$. Here and below, we have suppressed explicit time dependence $t$ for brevity. Importantly, the way the state changes with $\mathbf{A}$ directly produces the response to the physical electromagnetic fields $\mathbf{A}$~\cite{BernevigPhysRevLett.128.087002,Resta2022}, see explicit derivation in Supplementary Material {\bf (SM)} section 1~\cite{SM}\nocite{QNiu_1984, PhysRevLett.127.277202, Altland_Simons_2010, SCenhancement}. For e.g., the imaginary part of the quantum geometric tensor $\mathbf{G}_{\alpha \beta} (\mathbf{A}) = (\hbar^2/e^2){\rm Tr}[P(\partial^\alpha_{\mathbf{A}} P)(\partial^\beta_{\mathbf{A}} P)]$ is the Berry flux that determines the Hall conductivity \cite{RevModPhys.82.1959,Resta2022}; its real part is the quantum metric important for bounds of capacitance \cite{Raquel2024} and superfluidity \cite{Torma2015}. Here $P(\mathbf{A}) \equiv \ket{\Phi(\mathbf{A})}\bra{\Phi(\mathbf{A})}$ is a ground state projector. While reducing to the familiar $k$-dependent forms for noninteracting systems~\cite{Resta2022}, $\mathbf{G}_{\alpha \beta} (\mathbf{A})$ allows one to track physical responses in superconductors.

\begin{figure}[t]
    \centering
    \includegraphics[width=0.48\textwidth]{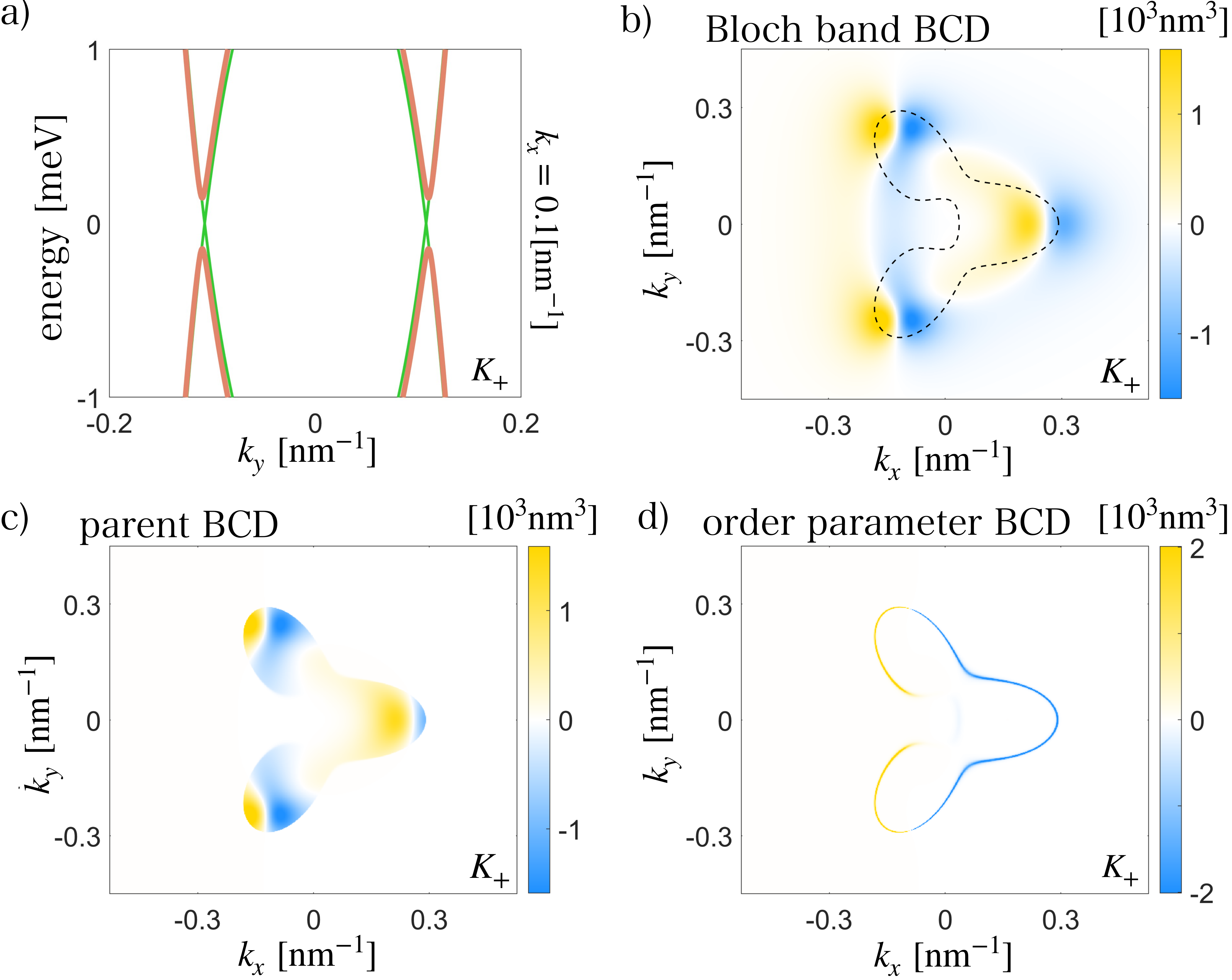}
    \caption{{\it $\mathcal{D}^{xyx}$ component of the Berry curvature dipole distribution in a noncentrosymmetric superconductor} {\bf a.} Fully gapped superconducting BdG band structure (red) for gated rhombohedral trilayer graphene (RTG) obtained from a self-consistent numerical solution of the mean-field Bogoliubov de Gennes Hamiltonian (see \textbf{SM} section 4~\cite{SM}). (green) Single-particle electron (and hole) dispersions $H_{\rm RTG}^{(0)} (\mathbf{k})$ ($-H_{\rm RTG}^{(0)} (-\mathbf{k})$) of the parent RTG Fermi surface. {\bf b.} Bloch state BCD distribution and the Fermi surface (black dashed line) of RTG. {\bf c.} Parent BCD distribution in a superconductor obtained by plotting integrand of the second term in Eq.~(\ref{SBCDIBA}). It resembles the single-particle Bloch BCD but is weighted by $|v_{\mathbf k,0}|^2$. {\bf d.} Order parameter BCD distribution obtained by plotting the integrand of the first term of Eq.~(\ref{SBCDIBA}) for RTG.  The distribution is sharply peaked close to the BdG band edges and is sensitive to the phase of the gap parameter. The color bar in panel {\bf d} is saturated; for the full scale see {\bf SM} section 6~\cite{SM}. In all panels, the superconducting Berry curvature dipole is shown for the $K_+$ valley. }
    \label{sBCDFIG1}
\end{figure}

Higher moments of quantum geometry can also be generated. Of special interest is the Berry curvature dipole:
\begin{equation}\label{generalBCD}
    \mathcal{D}^{\alpha\beta\gamma}=\frac{i\hbar^3}{e^3 V}
    \lim_{\mathbf{A}\rightarrow0}{\rm Tr}[P\partial^\alpha_{\mathbf{A}}(\partial^\beta_{\mathbf{A}} P)(\partial^\gamma_{\mathbf{A}} P)]- (\beta\leftrightarrow\gamma),
\end{equation}
where $V$ is the system volume (area in 2D) $\partial^\alpha_{\mathbf{A}}\equiv \partial/\partial A^\alpha$ and Greek indices denote Cartesian spatial components. Taking a metallic $\ket{ \Phi (\mathbf{A})} =\prod_{n,s,\mathbf{k}\leq \mathbf{k}_F}\hat{c}^{\dagger}_{\mathbf{k},n,s}(\mathbf{A})\ket{0}$ produces the BCD in a noninteracting metal as~\cite{SodemannFu,Resta2022}
\begin{equation}
\label{eq:DFS}
    \mathcal{D}^{\alpha\beta\gamma}_{\rm FS}=\sum_{n\in{\rm o}}\int_{\mathbf{k}}i\partial^\alpha_\mathbf{k}\left[(\partial_{\mathbf{k}}^\beta\bra{n,\mathbf{k}}) \partial_\mathbf{k}^\gamma\ket{n,\mathbf{k}}\right]- (\beta\leftrightarrow\gamma),
\end{equation} 
where $c^{\dag}_{\mathbf{k},n,s}(\mathbf{A})\ket{0} = e^{i\mathbf{k} \cdot {\mathbf{r}}} \ket{n, \mathbf{k} +\frac{e}{\hbar }\mathbf{A}}$ is the creation operator of the Bloch state with periodic part $\ket{n, \mathbf{k}}$ and $\int_{\mathbf{k}}\equiv \int d^d\mathbf{k}/(2\pi)^d$ with $\mathbf{k}$ the crystal momentum, $n$ the band index and $s$ the spin. Here and below, we have suppressed explicit $s$ dependence of the Bloch state for simplicity and ``${\rm o}$'' stands for occupied. Note, since BCD in Eq.~(\ref{eq:DFS}) is a full derivative, $\mathcal{D}^{\alpha\beta\gamma}_{\rm FS}$ vanishes for fully occupied bands \cite{SodemannFu}. Indeed, vanishing BCD in fully occupied bands is a property of noninteracting Bloch electrons~\cite{PhysRevB.98.155137}. As a result, $\mathcal{D}^{\alpha\beta\gamma}$ in Eq.~(\ref{generalBCD}) is often thought to manifest only for finite Fermi surfaces \cite{SodemannFu,OrtixPhysRevLett.123.196403}. 

In contrast, in superconductors where U(1) symmetry is broken, $\mathcal{D}^{\alpha\beta\gamma}$ can manifest without a Fermi surface and in a gapped phase (see Fig~\ref{sBCDFIG1}a). As a prime illustration, we consider a state:
\begin{equation}\label{SCGS}
    \ket{ {\rm SC }_{\mathbf{q}}(\mathbf{A})} \hspace{-1mm}=\hspace{-1mm}\prod_{\mathbf{k}}\hspace{-1mm}\left[ u_{\mathbf{k},\mathbf{q}}\hspace{-0.75mm}+\hspace{-0.75mm}v_{\mathbf{k},\mathbf{q}}\hat{c}^{\dagger}_{\mathbf{k}+\mathbf{q},\uparrow}(\mathbf{A})\hat{c}^{\dagger}_{-\mathbf{k}+\mathbf{q},\downarrow}(\mathbf{A})\right]\hspace{-1mm}\ket{0},
\end{equation}
that captures collective superconducting behavior in the thermodynamic limit~\cite{PhysRev.108.1175, PhysRev.112.1900, MattisLieb,parks1969superconductivity,DeGennes}. We have focused on the band that crosses the Fermi level and parameterized the superconducting state via the Cooper pair center-of-mass momentum $2\hbar \mathbf{q}$; $u_{\mathbf{k},\mathbf{q}}$ and $v_{\mathbf{k},\mathbf{q}}$ are coherence factors. 

To proceed we directly evaluate the superconducting Berry connection from Eq.~(\ref{SCGS}):
\begin{equation}
    \lim_{\mathbf{A}\rightarrow0}\!\frac{\hbar}{e}\langle {\rm SC}_{\mathbf{q}}(\mathbf{A}) | i\partial_{\mathbf{A}}^\gamma | {\rm SC}_{\mathbf{q}}(\mathbf{A}) \rangle\!=\!\sum_{\mathbf{k}} |v_{\mathbf{k},\mathbf{q}}|^2\, L_n^\gamma(\mathbf{k},\mathbf{q}),\label{MsBCon}
\end{equation}
where $L_n^\gamma (\mathbf{k}, \mathbf{q})$ is an emergent length scale (see below): 
\begin{equation}
    L_n^\gamma(\mathbf{k},\mathbf{q})=-\partial_{\mathbf{q}}^\gamma \arg v_{\mathbf{k},\mathbf{q}}+ r_n^\gamma(-\mathbf{k}+\mathbf{q})+ r_n^\gamma(\mathbf{k}+\mathbf{q}),\label{MgenL}
\end{equation}
with $r_n^\gamma(\mathbf{k})=\langle n,\mathbf{k}|\,i\partial_k^\gamma\,|n,\mathbf{k}\rangle$ the Bloch-state Berry connection. In obtaining the above, we used the convention where $u_{\mathbf{k},\mathbf{q}}$ is real~\cite{Tinkham} and we employed the fact that vector potential $e\mathbf{A}$ directly couples to the Cooper pair center-of-mass momentum $2\hbar\mathbf{q}$ (see \textbf{SM} section 2~\cite{SM} for discussion).

The quantity $\mathbf{L}_n(\mathbf{k},\mathbf{q})$ represents a novel ``Cooper pair shift.'' Distinct from the coherence length, $\mathbf{L}_{n} (\mathbf{k},\mathbf{q})$ in Eq.~(\ref{MgenL}) takes the familiar form of a shift vector \cite{ShiftLi}. Shift vectors arise from geometric phases accrued when an electron scatters off a potential \cite{Sinitsyn}, is photoexcited by light \cite{Ahn2022}, or when two fermions scatter off each other \cite{ThePesin, RoniIllan}. Here we find that when two electrons become bound to each other in a Cooper pair, a real-space ``Cooper pair shift'' emerges  [$\mathbf{L}_{n} (\mathbf{k},\mathbf{q})$] where pairing induces a collective displacement of the center-of-mass of the Cooper pair; as we will see below, it is sensitive to the structure of the order parameter. 

As we now explain, $\mathbf{L}_n(\mathbf{k}, \mathbf{q})$ directly determines the BCD of superconductors. To see this, we differentiate Eq.~(\ref{MsBCon}) with respect to $\partial^\beta_\mathbf{q}$ and $\partial^\alpha_\mathbf{q}$ and take the anti-symmetric part under exchange of $(\beta,\gamma)$ indices [this is equivalent to substituting Eq.(\ref{SCGS}) into Eq.(\ref{generalBCD})] to obtain the superconducting Berry curvature dipole (sBCD). Notably, both $L_n^\gamma$ and $|v_{\mathbf{k},\mathbf{q}}|^2$ depend on the center-of-mass momentum $2\hbar\mathbf{q}$. For time reversal symmetric systems, we find two distinct contributions to the sBCD ``order parameter BCD'' and a ``parent BCD'' (see also {\bf End Matter}):
\begin{multline}\label{SBCDIBA}
\mathcal{D}^{\alpha\beta\gamma}_{\rm sBCD}= \int_{\mathbf{k}}\overbrace{\left[ L_{n}^\gamma(\mathbf{k},\mathbf{q})\partial^\alpha_{\mathbf{q}}\partial^\beta_{\mathbf{q}}|v_{\mathbf{k,q}}|^2-(\beta\leftrightarrow\gamma)\right]_{\mathbf{q}\rightarrow0}}^{\text{order parameter BCD distribution}} \\+ 2\int_{\mathbf{k}}\underbrace{|v_{\mathbf{k,0}}|^2 \overbrace{(\partial^\alpha_{\mathbf{k}} \partial^\beta_{\mathbf{k}} r_{n}^\gamma (\mathbf{k})-\partial^\alpha_{\mathbf{k}} \partial^\gamma_{\mathbf{k}} r_{n}^\beta (\mathbf{k}))}^{\text{Bloch state~BCD distribution}}}_{\text{parent BCD distribution}}, 
\end{multline}
Here the parent BCD in Eq.~(\ref{SBCDIBA}) tracks the BCD distribution of the original Bloch states in the parent metal (Fig~\ref{sBCDFIG1}b); this contribution is a BCD inherited from the Bloch states while weighted by the electron density in the superconductor $|v_{\mathbf{k,0}}|^2$ resulting in the momentum distribution as shown in Fig.~\ref{sBCDFIG1}c. Notably, the weighting factor $|v_{\mathbf{k},0}|^2$ renders the parent BCD finite.

Critically, the first term of Eq.~(\ref{SBCDIBA}) reveals an order parameter BCD that has no analog in normal (non-superconducting) systems. It is a collective contribution that can be understood in terms of geometric phases accumulated by the order parameter as supercurrent flows (parameterized by $\mathbf{q}$). 

While the formulation above is expressed in terms of a generic BCS type wave function, for concreteness we evaluate these quantities within a mean-field Bogoliubov-de Gennes (BdG) framework. To this end, we estimate the coherence factors in the standard BdG description (see {\bf SM} section 2): $v_{\mathbf{k},\mathbf{q}} = -|v_{\mathbf{k},\mathbf{q}}|e^{i{\rm arg}\Delta_{\mathbf{k,q}}}$, where the order parameter is $\Delta_{\mathbf{k},\mathbf{q}} = \langle c_{\mathbf{k}+\mathbf{q},\uparrow}c_{-\mathbf{k}+\mathbf{q},\downarrow} \rangle = \bra{n,\mathbf{k+q}}\hat \Delta \ket{n,\mathbf{k-q}}$, with $\hat\Delta$ the multiband pairing potential operator. This yields $\mathbf{L}_{n}(\mathbf{k}, \mathbf{q})$ (and hence the order parameter BCD) that directly depends on the {\it phase} of the order parameter.

We find that $\mathbf{L}_{n} (\mathbf{k},\mathbf{q})$ is induced by the pairing potential with a nontrivial superconducting gap structure.  Indeed, in the limit $\mathbf{q}\rightarrow0$ it takes the geometrical form: 
\begin{equation}
\mathbf{L}_{n} (\mathbf{k},\mathbf{q})=-\boldsymbol{\partial}_{\mathbf{q}}{\rm arg}\left[\frac{\bra{n,\mathbf{k+q}}\hat{\Delta}\ket{n,\mathbf{k-q}}}{\langle{n,\mathbf{k+q}}\ket{n,\mathbf{k-q}}}\right].
\label{eq:shift}
\end{equation}
Notice that $\mathbf{L}_n(\mathbf{k}, \mathbf{q})$ emerges even when $\hat{\Delta}$ is s-wave and $\mathbf{k}$-independent; instead it captures the internal (sublattice) noncentrosymmetric structure of the pairing, see below for explicit example. In obtaining Eq.~(\ref{eq:shift}), we have noted $\langle{n,\mathbf{k+q}}\ket{n,\mathbf{k-q}} = \exp[2i\mathbf{r}_{n}\cdot \mathbf{q}+\mathcal{O}(q^2)]$ at small $\mathbf{q}$. For a structureless pairing potential $\hat\Delta = \Delta \mathbb{I}$, we find $\mathbf{L}_n(\mathbf{k}, \mathbf{q})$ {\it vanishes}. In contrast, when the pairing potential is non-centrosymmetric: e.g., when $\hat\Delta$ possesses a noncentrosymmetric orbital order, $\mathbf{L}_n(\mathbf{k}, \mathbf{q})$ is finite and contributes significantly to the total sBCD, see below. Critically, finite $\mathbf{L}_n(\mathbf{k}, \mathbf{q})$ is controlled by the (Cooper pair center-of-mass) $\mathbf{q}$-winding of the superconducting $\Delta_{\mathbf{k},\mathbf{q}}$ and encodes broken centrosymmetry in the superconducting pairing. 

Importantly, we find the order parameter BCD can dominate the total sBCD in noncentrosymmetric superconductors. Its distribution is sharply peaked for $\mathbf{k}$ close to the band edge of the lower Bogoliubov band (see Fig~\ref{sBCDFIG1}d and form of $|v_{\mathbf{k,q}}|^2$). Indeed, since the electrons and holes form gapped Bogoliubov bands (see Fig~\ref{sBCDFIG1}a), the order parameter BCD distribution traces out the Fermi surface. 

Even more striking is the case of inversion symmetric Bloch wave functions $\ket{n,\mathbf{k}} = \ket{n,\mathbf{-k}}$ in the parent metallic state. Inversion symmetry in the parent metallic state zeros the parent BCD. In contrast, inversion breaking in the pairing $\hat{\Delta}$ can still produce an order parameter BCD. This highlights the fact that order parameter BCD is a collective effect: sBCD can take on new collective features associated with superconducting order that can track its symmetry breaking. For example, this unusual property enables a superconducting {\it BCD proximity effect}: non-zero sBCD can be proximitized in a target centrosymmetric metal layer by a proximal noncentrosymmetric superconductor (see explicit discussion below). 

Notice that finite $\mathcal{D}^{\alpha\beta\gamma}$ requires both inversion (P) and composite inversion and time-reversal (PT) symmetries to be broken either by the lattice or the order parameter. In 2D systems rotational symmetries also constrain the sBCD~\cite{SodemannFu}, thus finite sBCD requires all in-plane rotational symmetries to be broken making mirror the largest point group symmetry compatible with sBCD in 2D. Interestingly, this makes $\mathcal{D}^{\alpha\beta\gamma}$ highly sensitive to broken rotational symmetry; it can even be used as a diagnostic of spontaneous ordering (e.g., from nematicity).

\textit{sBCD Hall inertia and nondissipative responses}. To illustrate sBCD responses, we focus on dynamics of a driven superconductor by slowly varying modulations, $0<\hbar\omega \ll \Delta$. Here we parameterize the driving through a time-varying center-of-mass momentum of the Cooper pairs, $\mathbf{q}\rightarrow \mathbf{q}(t)$. Note that $\dot{\mathbf{q}}(t) =- e\mathbf{E}(t)/\hbar$ is produced by an oscillating electric field. Its impact can be readily appreciated by examining how the velocity density of superconductor changes with $\delta \mathbf{q}(t)$:  
\begin{equation}
\delta v^\gamma(t) = \frac{\partial^2 \mathcal{E}(\mathbf{q})}{\partial q^\alpha \partial q^\gamma}  \frac{ q^\alpha(t)}{V\hbar} +   q^\alpha (t)\mathcal{D}_{\rm sBCD}^{\alpha \beta \gamma} \dot{q}^\beta (t) , 
\label{eq:nonnewtonian}
\end{equation}
where $\hat{H}_{\mathbf{q}}\ket{{\rm SC}_{\mathbf{q}}} = \mathcal{E} (\mathbf{q})\ket{{\rm SC}_{\mathbf{q}}}$ with $\hat{H}_{\mathbf{q}}$ the Hamiltonian of the superconductor and $\mathcal{E} (\mathbf{q})$ its ground state energy. The first term in Eq.~(\ref{eq:nonnewtonian}) directly mirrors a classical Newtonian velocity-momentum relationship with $\partial^2 \mathcal{E}(\mathbf{q})/\partial q^\alpha \partial q^\gamma$ an inverse mass, while the second originates from a non-Newtonian acceleration induced by quantum geometry \cite{matsyshyn2019nonlinear}.

\begin{figure*}
\centering
\includegraphics[width=0.98\textwidth]{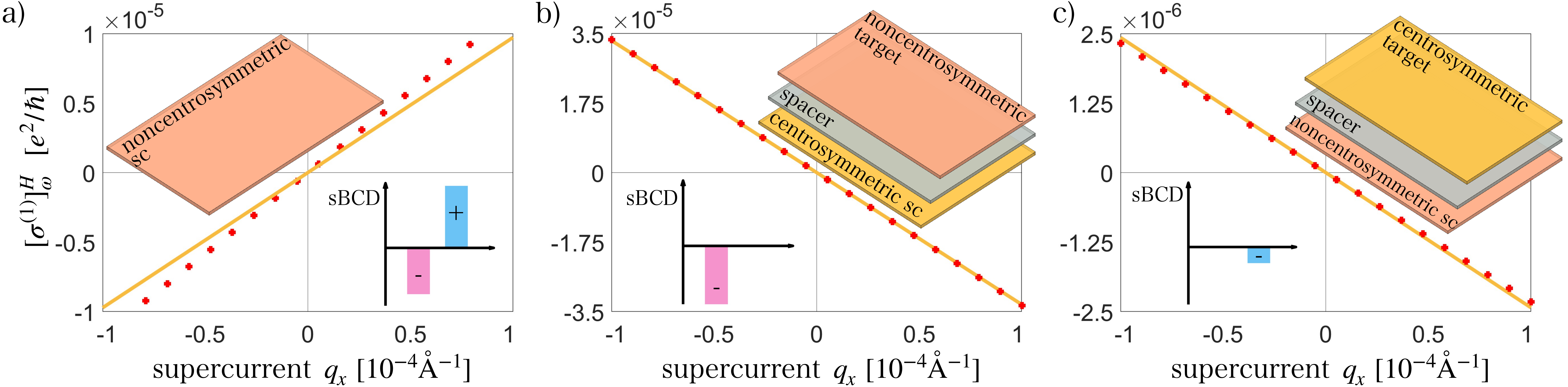}
\caption{{\it Superconducting Berry curvature dipole  and nondissipative supercurrent-induced dynamical Hall conductivity.} Dynamical Hall conductivity as a function of supercurrent (Cooper pair center-of-mass momentum $2\hbar q_x$) of a target material comparing prediction of Eq.(\ref{eq:conductivities}) (orange line) and full multiband result (red dots, see {\bf SM} section 5~\cite{SM}) in three setups: {\bf a.} non-centrosymmetric superconductor with a noncentrosymmetric orbital structure for the superconductor pairing $\hat\Delta = [\Delta_{A},0;0,\Delta_{B}]$ {\bf b.} non-centrosymmetric target layer with a centrosymmetric proximity induced pairing structure $\hat\Delta_0^{\rm proximity} = [\Delta_{0},0;0,\Delta_{0}]$ {\bf c.} centrosymmetric target layer with a noncentrosymmetric proximity effect: pairing possesses a non-trivial orbital structure $\hat\Delta^{\rm proximity} = [\Delta^{\rm proximity}_{A},0;0,\Delta^{\rm proximity}_{B}]$. Bottom inset illustrates contributions from order parameter (blue) and parent (pink) components to the total sBCD. See {\bf SM} section 3-4~\cite{SM} for a full discussion of the RTG parameters and the self-consistent solution of the BdG multiband pairing potential.}
\label{sBCDFIG2}
\end{figure*}

In the same fashion as the BCD in a metal \cite{SodemannFu, matsyshyn2019nonlinear,PhysRevB.107.125151}, the collective $\mathcal{D}_{\rm sBCD}$ produces a non-Newtonian acceleration that is responsible for electromagnetic second-order nonlinearities $j_{(2)}^\gamma(t) = \int_{\omega,\omega_1} e^{-i\omega t}  [\sigma^{(2)}]^{\gamma\alpha\beta}_{\omega,\omega_1} E^\alpha_{\omega_1} E^\beta_{\omega-\omega_1}$ where $\int_\omega \equiv \int_{-\infty}^\infty d\omega/2\pi$ and $\mathbf{E}(t) =\int_\omega e^{-i\omega t}\mathbf{E}_\omega$. Indeed, collecting terms second order in electric field in Eq.~(\ref{eq:nonnewtonian}) yields a pure imaginary $[\sigma^{(2)}]^{\gamma\alpha\beta}_{\omega,\omega_1} = ie^3\mathcal{D}^{\alpha\beta\gamma}_{\rm sBCD}/(\hbar^2\omega_1)$ \cite{Yanase1,Yanase2} in a superconductor that only depends on the frequency of the drive and $\mathcal{D}_{\rm sBCD}$, see also {\bf SM} section 5~\cite{SM} for a short discussion and estimate. 

Importantly, notice that $\mathcal{D}^{\alpha\beta\gamma}_{\rm sBCD}$ includes both parent (similar to Bloch band BCD) and order parameter (collective) contributions, as highlighted in Eq.~(\ref{SBCDIBA}). As a result, order parameter BCD provides an additional contribution to second-order nonlinearities that cannot be found in a purely noninteracting parent metallic state. This mirrors the situation for the superfluid weight [first term in Eq.~(\ref{eq:nonnewtonian})] which includes both a noninteracting Drude contribution \cite{Tinkham} and collective quantum geometric contributions \cite{Torma2015,BernevigPhysRevLett.128.087002,Lapa2019}.  

We now focus on a new effect that sBCD can produce in a superconductor distinct from that of a metal. To see this, we observe that a DC center-of-mass momentum in a superconductor $2\hbar\mathbf{q}_{\rm sc}$ does not relax. Writing $\mathbf{q} (t) = \mathbf{q}_{\rm sc} + \mathbf{q}_{\rm ac} {\rm cos}(\omega t)$ and isolating the linear Hall response in Eq.~(\ref{eq:nonnewtonian}) we find the sBCD produces a dynamical linear Hall effect: 
\begin{equation}
    \sigma^{(1)}_H (\omega) \hspace{-1mm}=\hspace{-0.5mm} 
    e[\hat{\rho}^{-1}\mathbf{j}_0]^{\alpha} \mathcal{D}^{\alpha xy}_{\rm sBCD}, \quad \mathbf{q}_{\rm sc} =\frac{\hbar}{e} \hat{\rho}^{-1}\mathbf{j}_0 
    \label{eq:conductivities}
\end{equation}
where $j_0$ is DC supercurrent density and $\rho^{\alpha \gamma}=  \partial_\mathbf{q}^\gamma\partial^\alpha_{\mathbf{q}} \mathcal{E}(\mathbf{q})/V$ is the superfluid weight. Importantly, $\sigma^{(1)}_H$ is a {\it nondissipative} dynamical linear Hall conductivity, induced by running a supercurrent. This contrasts with metals where current-induced Hall effects are necessarily dissipative~\cite{PhysRevB.99.155404} due to extrinsic scattering across a gapless Fermi surface. In our case, the superconducting gap quenches extrinsic disorder scattering for frequencies below the gap (states at the Fermi surface are gapped out). This renders superconductors a unique venue to realize intrinsic BCD responses ``under the gap''. This mirrors the dichotomy of the magnetoelectric effect \cite{Levitov} that can be nondissipative for superconductors but dissipative for metals. 

The induced $\sigma^{(1)}_H$ in Eq.~(\ref{eq:conductivities}) and Eq.~(\ref{eq:nonnewtonian}) provides a natural way to physically interpret order parameter BCD: it represents an emergent Hall inertia that develops when electrons Cooper pair. Much as interactions enable to modify inverse mass, they also induce a transverse inertia in the presence of a supercurrent. 

\textit{sBCD in noncentrosymmetric superconducting heterostructures}. We now illustrate sBCD in a two dimensional noncentrosymmetric material within a mean-field BdG framework (see e.g., {\bf SM} sections 2-4~\cite{SM} and Refs. \cite{Yanase1,Yanase2}). We will focus on rhombohedral trilayer graphene (RTG) where centrosymmetry can be readily turned on and off by an external gate. This broken centrosymmetry can be described by an effective low-energy bare Hamiltonian for RTG \cite{Zhang_2010,Jung_2013}: $\hat{H}_{\rm RTG}^{(0)} = \mathbf{d}(\mathbf{k}) \cdot \boldsymbol{\tau} + \delta g(\mathbf{k}) \tau_z$ where $\boldsymbol{\tau}$ are Pauli matrices describing A/B sublattices, $\delta$ describes the (external gate-tunable) inversion symmetry breaking and $\mathbf{d}(\mathbf{p})$ captures the pristine centrosymmetric behavior of RTG when $\delta =0$, see full description and microscopic description of $\hat{H}_{\rm RTG}^{(0)}$ and the functional form of $\mathbf{d}(\mathbf{k}),g(\mathbf{k})$ in {\bf SM} section 3~\cite{SM}.  

Noncentrosymmetric RTG has been observed to host an intrinsic superconducting state \cite{Zhou2021}. In what follows, we model the superconducting state via a BdG Hamiltonian $\hat{H}= \frac{1}{2}\sum_{\mathbf{k}}\Psi^\dagger_{\mathbf{k,q}} \hat{H}^{\rm BdG}_{\mathbf{k,q}}[\mathbf{A}(t)]\Psi_{\mathbf{k,q}}$ in the Bloch-Nambu basis $\Psi_{\mathbf{k,q}}=(\vec{c}_{\mathbf{k}+\mathbf{q}},\vec{c}^{~\dagger}_{-\mathbf{k}+\mathbf{q}})^{\mathrm{T}}$ (here we use $\vec{c}$ to denote a multicomponent pseudospin). We adopt a local attractive density-density interaction potential~\cite{LevitovSC} and self-consistently solved $\Delta_{\mathbf{q},ab} = -U\sum_{\mathbf{k}}\langle c_{-\mathbf{k}+\mathbf{q},a}c_{\mathbf{k}+\mathbf{q},b}\rangle$ \cite{DeGennes} (for more details see \textbf{SM} section 4 \cite{SM}). Here $a,b$ runs over the $\{A,B\}$ sublattice degree of freedom yielding a multiparameter superconducting gap matrix $\hat\Delta$. In RTG $\hat\Delta = [\Delta_{A},0;0,\Delta_{B}]$, breaks inversion symmetry  for $\Delta_A \neq \Delta_B$ \cite{LevitovSC, SM}. Using this self-consistent solution, we numerically plot the distribution for the parent and order parameter BCD components of $\mathcal{D}^{xyx}_{\rm sBCD}$ in Fig.~\ref{sBCDFIG1}c,d respectively. While parent BCD distribution is diffuse and occurs predominantly for $\mathbf{k}$ values inside the metallic Fermi surface of RTG, order parameter BCD distribution is sharply peaked. It is useful to note that for RTG, $x = (\Delta_A-\Delta_B)/(\Delta_A+\Delta_B)$ controls the strength of the Cooper pair shift $L_n (\mathbf{k},\mathbf{q})$ for small $x$; it vanishes when $\Delta_A= \Delta_B$. 

Importantly, in this uniform bulk system, parent and order parameter BCDs are comparable in magnitude but have opposite signs; the net effect of sBCD arises from their competition. To see this, we plot the dynamical Hall conductivity in Fig.~\ref{sBCDFIG2}a (orange lines). The  relative contributions of order parameter and parent BCDs are depicted in the lower inset with order parameter outweighing that of parent BCD. We check this by performing a multiband calculation \cite{Yanase1,Yanase2} of the linear conductivity in the superconducting state (red dots) using a quantum Louiville approach, see {\bf SM} section 5~\cite{SM}. This full treatment reproduces the discussed results, with full multiband $\sigma_H^{(1)}$ agreeing well with Eq.~(\ref{eq:conductivities}). 

To exhibit the importance of the A-B structure of the order parameter in sBCD, we next examine a noncentrosymmetric target metallic layer proximitized by a centrosymmetric superconductor. For simplicity, we use noncentrosymmetric $\hat{H}^{(0)}_{\rm RTG}$ in its metallic state as the target layer. Because the superconductor substrate (see schematic inset in Fig.~\ref{sBCDFIG2}b) does not break inversion symmetry, this proximity effect can be modeled with a centrosymmetric pairing $\hat\Delta_0^{\rm proximity} = [\Delta_{0},0;0,\Delta_{0}]$ producing $\sigma_H^{(1)}$ in Fig.~\ref{sBCDFIG2}b. In this case, order parameter BCD vanishes (see lower inset). Nevertheless, $\sigma_H^{(1)}$ persist and are controlled by parent BCD.  

For the reverse situation: a centrosymmetric target metallic layer is proximitized by a non-centrosymmetric superconductor. Indeed, non-centrosymmetric proximity effect into centrosymemtric graphene systems have been viably proposed~\cite{PhysRevB.99.235404}; recent experiments have reported preliminary evidence that superconducting correlations from monolayer NbSe$_2$ can be proximitized in graphene systems \cite{Naritsuka2025}. Here we choose $\hat{H}^{(0)}_{\rm RTG}$ with $\delta = 0$ to model the centrosymmetric target layer; due to its centrosymmetry its metallic state should exhibit a zero current induced dynamical Hall conductivity. Similarly, Bloch state BCD vanishes. However, modeling the proximity effect from the non-centrosymmetric superconductor as $\hat\Delta^{\rm proximity} = [\Delta^{\rm proximity}_{A},0;0,\Delta^{\rm proximity}_{B}]$ with $\Delta^{\rm proximity}_{A} \neq \Delta^{\rm proximity}_{B}$ produces an order parameter BCD. Even as parent BCD vanishes, the order parameter BCD allows finite $\sigma_H^{(1)}$ to develop (see Fig. 2c). This demonstrates a type of superconducting BCD proximity effect made possible by the order parameter BCD. 

The sBCD is {\it collective} in nature and sharply departs from the familiar BCD for noninteracting electrons. In particular, order parameter BCD demonstrates how quantum geometry is transformed in superconducting quantum states. Interestingly, order parameter BCD depends on the $\mathbf{q}$-gradient of the phase of $\Delta_{\mathbf{k},\mathbf{q}}$, rendering sBCD responses a novel probe of the phase structure of  the order parameter. SBCD produces {\it nondissipative responses} with operating frequencies below the superconducting gap distinct from their metallic counterparts rendering them an attractive platform to exploit quantum geometric nonlinearities for superconducting circuits and components~\cite{Fatemi.21.064029}. In this regard, we expect sBCD transverse responses (e.g., $\sigma_H^{(1)}$) are best probed at high frequencies (e.g., microwave frequencies) where superconducting longitudinal responses are suppressed. While we have focused on responses deep in the superconducting state, fluctuations close to the transition temperature may play a role enriching nonlinear response and enhancing their magnitude~\cite{dong2025enhanced}; indeed, the role of fluctuations on quantum geometry particularly for 2D systems, as well as the the interplay of order parameter vs parent Bloch band contributions to quantum geometry are important areas for future work. Looking forward, we anticipate that the fundamental underlying mechanism behind sBCD can be applied to a wide range of materials as well as responses beyond superconductors, e.g., polarization responses of excitonic states~\cite{resta2024geometrical, yang2025correlatedquantumshiftvector}.

\textit{Acknowledgements}. We gratefully acknowledge useful conversations with Valla Fatemi, Roshan Krishna Kumar, and Inti Sodemann, Xu Yang. This research was supported by the Singapore Ministry of Education (MOE) Academic Research Fund Tier 2 grant (MOE-T2EP50222-0011) (JCWS) and Tier 3 grant (MOE-MOET32023-0003) “Quantum Geometric Advantage” (JCWS). GV was supported by the Ministry of Education, Singapore, under its Research Centre of Excellence award to the Institute for Functional Intelligent Materials (I-FIM, project No. EDUNC-33-18-279-V12).

\bibliography{BCDsupercond}
\clearpage
\onecolumngrid
\section*{\large End Matter}
\twocolumngrid
{\textbf{Superconducting BCD from BCS wave function}}. In this End Matter, we demonstrate how to evaluate explicit expressions for the superconducting Berry curvature dipole discussed earlier. To do so, we first note that the projector 
$P(\mathbf{A})=\ket{\Phi(\mathbf{A})}\bra{\Phi(\mathbf{A})}$ satisfies 
\begin{equation}\label{projdef_new}
    \hspace{-2mm}\Big[{\rm Tr}[P\partial^\alpha_{\mathbf{A}}(\partial^\beta_{\mathbf{A}} P)(\partial^\gamma_{\mathbf{A}} P)] \Big]^{-}
    \hspace{-3mm}=\Big[\partial^\alpha_{\mathbf{A}} \bra{ \partial^\beta_{\mathbf{A}} \Phi (\mathbf{A})}{\partial^\gamma_{\mathbf{A}} \Phi (\mathbf{A})}\rangle\Big]^{-}\hspace{-2mm},
\end{equation}
where $[\mathcal{O}^{\alpha \beta \gamma}]^{-} = \mathcal{O}^{\alpha \beta \gamma} - \mathcal{O}^{\alpha \gamma \beta}$. Eq.~(\ref{projdef_new}) can be directly used to evaluate Berry curvature dipole in Eq.~(\ref{generalBCD}) for an \textbf{A} dependent quantum state $\ket{\Phi(\mathbf{A})}$. We first illustrate this procedure for the Fermi sea. Thereafter, we analyze the BCS wave function.

{\it Fermi-sea.} Writing the Fermi-sea as $\ket{ {\rm FS} (\mathbf{A})} =\prod_{s,{n \mathbf{k}}\in{\rm o}}\hat{c}^{\dagger}_{\mathbf{k},n,s}(\mathbf{A})\ket{0}$ and taking a derivative produces 
\begin{multline}\label{manybodyBC}
    \langle {\rm FS} (\mathbf{A})|\partial_\mathbf{A}^\gamma\ket{ {\rm FS} (\mathbf{A})} =\sum_{i}\bra{0}\hat{c}_{\mathbf{k}_i}(\mathbf{A})[\partial_\mathbf{A}^\gamma\hat{c}^{\dagger}_{\mathbf{k}_i}(\mathbf{A})]\ket{0} \\= \sum_{\{ n,\mathbf{k}\} \in{\rm o}} \bra{n,\mathbf{k}+\frac{e}{\hbar}\mathbf{A}} \partial_\mathbf{A}^\gamma\ket{n,\mathbf{k}+\frac{e}{\hbar}\mathbf{A}}. 
\end{multline} 
where $i$ sums across $\hat{c}_{\mathbf{k}_i}^\dagger$ in the Fermi sea for which $\partial_\mathbf{A}$ acts on; all other $\mathbf{k}$ in $\ket{ {\rm FS} (\mathbf{A})}$ (``un-differentiated pairs'') collapse after anticommutation with their partner $\hat{c}_{\mathbf{k}}$. By differentiating Eq.~(\ref{manybodyBC}) with respect to $\partial_\mathbf{A}^\beta$ and $\partial_\mathbf{A}^\alpha$, and isolating the antisymmetric part of the result in $\gamma,\beta$ we reproduce the Berry curvature dipole~\cite{Resta2022} in Eq.~(\ref{eq:DFS}). 

\textit{Superconducting state.} This same procedure can be directly applied to the BCS state~\cite{parks1969superconductivity}: $\ket{ {\rm SC }_{\mathbf{q}}(\mathbf{A})} \hspace{-1mm}=\hspace{-1mm}\prod_{\mathbf{k}}\hspace{-1mm}\left[ u_{\mathbf{k},\mathbf{q}}\hspace{-0.75mm}+\hspace{-0.75mm}v_{\mathbf{k},\mathbf{q}}\hat{c}^{\dagger}_{\mathbf{k}+\mathbf{q},\uparrow}(\mathbf{A})\hat{c}^{\dagger}_{-\mathbf{k}+\mathbf{q},\downarrow}(\mathbf{A})\right]\hspace{-1mm}\ket{0}$ with band index suppressed for brevity and coherence factors $u_{\mathbf{k},\mathbf{q}}$ and $v_{\mathbf{k},\mathbf{q}}$. Enumerating the momenta $\mathbf{k}\in (\mathbf{k}_1,\mathbf{k}_2, \ldots,\mathbf{k}_N)$, we evaluate
\begin{align}
&\langle { {\rm SC }(\mathbf{A})}\ket{\partial_\mathbf{A}^\alpha {\rm SC }(\mathbf{A})} =\nonumber\\&\sum_i\bra{0}\left[ u^*_{\mathbf{k_1},\mathbf{q}}(\mathbf{A})+v^*_{\mathbf{k_1},\mathbf{q}}(\mathbf{A})\hat{c}_{-\mathbf{k_1}+\mathbf{q},\downarrow}(\mathbf{A})\hat{c}_{\mathbf{k_1}+\mathbf{q},\uparrow}(\mathbf{A})\right]\ldots\nonumber\\&\times\left[ u^*_{\mathbf{k_N},\mathbf{q}}(\mathbf{A})+v^*_{\mathbf{k_N},\mathbf{q}}(\mathbf{A})\hat{c}_{-\mathbf{k_N}+\mathbf{q},\downarrow}(\mathbf{A})\hat{c}_{\mathbf{k_N}+\mathbf{q},\uparrow}(\mathbf{A})\right]\nonumber\\&\times[ u_{\mathbf{k_N},\mathbf{q}}(\mathbf{A})+v_{\mathbf{k_N},\mathbf{q}}(\mathbf{A})\hat{c}^\dagger_{\mathbf{k_N}+\mathbf{q},\uparrow}(\mathbf{A})\hat{c}^\dagger_{\mathbf{-k_N}+\mathbf{q},\downarrow}(\mathbf{A})]\times\ldots\nonumber\\&\times\Big\{ \partial_\mathbf{A}^\alpha u_{\mathbf{k}_i,\mathbf{q}}(\mathbf{A})+( \partial_\mathbf{A}^\alpha v_{\mathbf{k}_i,\mathbf{q}}(\mathbf{A}))\hat{c}^\dagger_{\mathbf{k}_i+\mathbf{q},\uparrow}(\mathbf{A})\hat{c}^\dagger_{-\mathbf{k}_i+\mathbf{q},\downarrow}(\mathbf{A}) \nonumber\\&\quad+v_{\mathbf{k_i},\mathbf{q}}(\mathbf{A}) \partial_\mathbf{A}^\alpha[\hat{c}^\dagger_{\mathbf{k}_i+\mathbf{q},\uparrow}(\mathbf{A})\hat{c}^\dagger_{-\mathbf{k}_i+\mathbf{q},\downarrow}(\mathbf{A})] \Big\}\ldots\ket{0},
\end{align}
where the summation over $i$ means that derivative $\partial_{\mathbf{A}}^\alpha$ acts on the contents of each set of large curly braces $\{ \cdots \}$; we sum all such derivatives. Collapsing all the undifferentiated pairs of operators of $\hat{c}^{\dagger}(\mathbf{A})$ and $\hat{c}(\mathbf{A})$ yields
\begin{multline}
    \langle { {\rm SC }(\mathbf{A})}\ket{\partial_\mathbf{A}^\alpha {\rm SC }(\mathbf{A})} \hspace{-0.75mm}=\hspace{-0.75mm}\\\sum_i\bigg\{u^*_{\mathbf{k}_i,\mathbf{q}}(\mathbf{A})\partial_\mathbf{A}^\alpha u_{\mathbf{k}_i,\mathbf{q}}(\mathbf{A})+v^*_{\mathbf{k}_i,\mathbf{q}}(\mathbf{A})\partial_\mathbf{A}^\alpha v_{\mathbf{k}_i,\mathbf{q}}(\mathbf{A}) +\\+|v_{\mathbf{k}_i,\mathbf{q}}(\mathbf{A})|^2\bra{0}\big( \hat{c}_{-\mathbf{k}_i+\mathbf{q},\downarrow}(\mathbf{A})\hat{c}_{\mathbf{k}_i+\mathbf{q},\uparrow}(\mathbf{A})\times\\\times\partial_\mathbf{A}^\alpha[\hat{c}^\dagger_{\mathbf{k}_i+\mathbf{q},\uparrow}(\mathbf{A})\hat{c}^\dagger_{-\mathbf{k}_i+\mathbf{q},\downarrow}(\mathbf{A})] \big)\ket{0}\bigg\}.
    \label{eq:SCAv2}
\end{multline}
To make progress, we notice that $\mathbf{A}$ directly couples with the center-of-mass momentum of the Cooper pairs $\mathbf{q}$ because electrons forming a Cooper pair have the same charge. As a result, derivatives with respect to $\frac{e}{\hbar}\mathbf{A}$ are equivalent to derivatives with respect to $\mathbf{q}$. Writing the Bloch Berry connection of the states as $r_n^\alpha(\mathbf{k}) \equiv \bra{n,\mathbf{k}}i\partial_\mathbf{k}^\alpha \ket{n,\mathbf{k}}$ and recalling the normalization condition for coherence factors $|u_{\mathbf{k},\mathbf{q}}(\mathbf{A})|^2+|v_{\mathbf{k},\mathbf{q}}(\mathbf{A})|^2=1$, we obtain a simplified superconducting Berry connection in Eq.(\ref{MsBCon}).

Plugging Eq.(\ref{MsBCon}) into Eq.(\ref{projdef_new}), we find the general expression for the sBCD as
\begin{multline}\label{genSBCD}
\mathcal{D}^{\alpha\beta\gamma}_{\rm sBCD}\hspace{-0.75mm}=\hspace{-0.75mm} \lim_{\mathbf{q}\rightarrow0}\sum_{\mathbf{k}}\hspace{-0.5mm} \hspace{-0.5mm}\bigg[(\partial^\beta_\mathbf{q}|v_{\mathbf{k},\mathbf{q}}|^2) \partial_\mathbf{q}^\alpha L^\gamma_n(\mathbf{k,q})\\+\frac{1}{2}(\partial_\mathbf{q}^\alpha\partial^\beta_\mathbf{q}|v_{\mathbf{k},\mathbf{q}}|^2) L^\gamma_n(\mathbf{k,q})+\frac{1}{2}|v_{\mathbf{k},\mathbf{q}}|^2 \partial_\mathbf{q}^\alpha\partial^\beta_\mathbf{q} L^\gamma_n(\mathbf{k,q})\\+(\alpha\leftrightarrow\beta)\bigg]-\hspace{-0.5mm}(\gamma\leftrightarrow\beta). 
\end{multline}
Notice that the term in the first line vanishes under time-reversal symmetry (TRS). In contrast, the terms in the second line persist even with TRS and are the components of the sBCD reported in this work: the first term in the second line produces the order parameter sBCD, while the second term in the second line produces the parent sBCD. For TRS, Eq.~(\ref{genSBCD}) becomes Eq.~(\ref{SBCDIBA}).  

\clearpage
\appendix
\onecolumngrid
\setcounter{equation}{0}
\setcounter{page}{1}
\setcounter{figure}{0}
\renewcommand{\theequation}{S-\arabic{equation}}
\renewcommand{\thefigure}{S-\arabic{figure}}
\renewcommand{\thetable}{S-\Roman{table}}
\titleformat{\section}[block]{\normalfont\normalsize\bfseries\centering}{Section \Alph{section}:}{0.3em}{}
\section*{\large Supplemental Material for ``Superconducting Berry Curvature Dipole''}
\subsection*{1. Adiabatic response of an electronic system.}
In this section, we study an adiabatic evolution of a system electromagnetically driven with a vector potential $\mathbf{A}(t)$. Importantly, as we will see below, the way wave functions vary with an applied vector potential $\mathbf{A}(t)$ is fundamental to physical responses (such as the Hall current) as it captures how an electronic system couples with applied electromagnetic fields. As an example, consider an adiabatic Hamiltonian $\hat H[\mathbf{A}]$ with the instantaneous ground state $\ket{\Phi(\mathbf{A})}$ and energy $\mathcal{E}(\mathbf{A})$. The instantaneous current density then can be written as $\mathbf{j} = e \mathbf{v}$, where the velocity density has two components: expectation of the current operator and the Niu-Thouless adiabatic current, written as \cite{Resta2022}:
\begin{equation} \label{deltavi}
\begin{aligned}
    e{v}^\gamma &= \frac{1}{V} \bra{\Phi(\mathbf{A})} (\partial^\gamma_\mathbf{A}\hat H[\mathbf{A}])\ket{\Phi(\mathbf{A})} -\frac{i\hbar}{V}\left(\bra{\partial^\gamma_\mathbf{A}\Phi(\mathbf{A})}\partial_t\Phi(\mathbf{A})\rangle - \bra{\partial_t\Phi(\mathbf{A})}\partial^\gamma_\mathbf{A}\Phi(\mathbf{A})\rangle\right)\\&= \frac{1}{V} \partial^\gamma_\mathbf{A}\mathcal{E}(\mathbf{A}) +\frac{i\hbar}{V}\dot{A}^\beta\left(  \bra{\partial_{\bf A}^\beta\Phi(\mathbf{A})}\partial^\gamma_\mathbf{A}\Phi(\mathbf{A})\rangle-\bra{\partial^\gamma_\mathbf{A}\Phi(\mathbf{A})}\partial_{\bf A}^\beta\Phi(\mathbf{A})\rangle\right), 
\end{aligned}
\end{equation}
where the first term is a group velocity and the second term a Berry flux. Notice that in this formulation no explicit $k$ dependence appears since $\ket{\Phi(\mathbf{A})}$ is a full state. Recalling that $\partial_t A(t) = -E(t)$ the electric field, the second term naturally produces the Hall conductivity \cite{QNiu_1984}.

We note that perhaps the most widely known form of  Eq.(\ref{deltavi}) is its single-particle noninteracting form \cite{RevModPhys.82.1959}, when the vector potential boosts the by the wave-vector $\mathbf{k}(t) = \mathbf{k}(0) +\frac{e}{\hbar}\mathbf{A}(t)$. In this noninteracting limit the second term of Eq.~(\ref{deltavi}) reduces to the sum of individual single-particle Berry curvature across $\mathbf{k}$. 

However, as shown in Eq.~(\ref{deltavi}), the response can be tracked by examining how the wave function changes with $\mathbf{A}(t)$. Indeed, for distinct states, the action of $\mathbf{A}$ can be distinct depending on what the collective coordinates are. For example, for the BCS groundstate formed of Cooper pairs, the vector potential couples with the center-of-mass momentum of the Cooper pair $2\hbar\mathbf{q}$. In contrast, for excitonic states formed of an electron-hole pair, the vector potential couples instead to the relative momentum between the electron and hole (since electron and hole have opposite charge) \cite{yang2025correlatedquantumshiftvector}. 

Considering the adiabatic evolution of all the terms in the Eq~(\ref{deltavi}), we find that change in velocity $\delta\mathbf{v}(\mathbf{A})=\mathbf{v}(\mathbf{A}) -\mathbf{v}(0)$ has two contributions:
\begin{equation}
    e\delta{v}^\gamma = \frac{1}{V} A^\alpha \partial^\gamma_\mathbf{A}\partial^\alpha_\mathbf{A}\mathcal{E}(\mathbf{A}) +\frac{i\hbar}{V}A^\alpha\dot{A}^\beta\partial_{\mathbf{A}}^\alpha\left(  \bra{\partial_{\bf A}^\beta\Phi(\mathbf{A})}\partial^\gamma_\mathbf{A}\Phi(\mathbf{A})\rangle-\bra{\partial^\gamma_\mathbf{A}\Phi(\mathbf{A})}\partial_{\bf A}^\beta\Phi(\mathbf{A})\rangle\right),
\end{equation}
where the Hall component is controlled by the BCD as discussed in the main text and End Matter. 

Noting that in superconductors the vector potential directly couples with the center-of-mass momentum of the Cooper pairs $2\hbar\mathbf{q}$, we can write $\partial_{\mathbf{A}} = \frac{e}{\hbar}\partial_{\mathbf{q}}$ and $\dot{\mathbf{q}}(t) = \frac{e}{\hbar}\dot{\mathbf{A}}(t)$ (see also below). This shows that change in velocity with a momentum boost of the Cooper pairs can be separated in two contribution: superfluid weight and the superconducting BCD. As a result, we obtain Eq.~(\ref{eq:nonnewtonian}) of the main text. 

\subsection*{2. Bogoliubov de Gennes (BdG) equation and the isolated band approximation.} In the following we review the BdG equations starting from a real-space treatment and discuss the isolated band approximation. The BdG equations extend the Bardeen-Cooper-Schrieffer theory to non-homogeneous systems. These are coupled equations for the complex amplitudes $\ket{\mathcal{U}(r)}$ and $\ket{\mathcal{V}(r)}$ in the presence of a finite super-current, where the phase of the gap function becomes a linear function of the center-of-mass coordinate:
\begin{align}
\label{eq:realspacebdg}
    &\left\{\frac{1}{2}[-i\boldsymbol{\nabla} + \frac{e}{\hbar}\mathbf{A}(\mathbf{r,t})]^2\delta_{ss'} + V_{ss'} (\mathbf{r}) - \mu\delta_{ss'}\right\}{\mathcal{U}_{s'}(\mathbf{r})} +\int d\mathbf{r}'e^{+i\mathbf{q}\cdot(\mathbf{r}+\mathbf{r}')}\Delta_{ss'}(\mathbf{r}, \mathbf{r}'){\mathcal{V}_{s'}(\mathbf{r}')} = E{\mathcal{U}_{s}(\mathbf{r})},\\-&\left\{\frac{1}{2}[+i\boldsymbol{\nabla} + \frac{e}{\hbar}\mathbf{A}(\mathbf{r,t})]^2\delta_{ss'} + V^*_{ss'} (\mathbf{r}) - \mu\delta_{ss'}\right\}{\mathcal{V}_{s'}(\mathbf{r})} +\int d\mathbf{r}'e^{-i\mathbf{q}\cdot(\mathbf{r}+\mathbf{r}')}[\Delta_{ss'}(\mathbf{r}, \mathbf{r}')]^\dagger{\mathcal{U}_{s'}(\mathbf{r}')} = E{\mathcal{V}_{s}(\mathbf{r})}.\label{eq:realspacebdg2}
\end{align}
An important property of this equation is its \textit{gauge covariance}. Namely if $\ket{u(r)}$ and $\ket{v(r)}$ are solutions with potentials $\hat V(\mathbf{r}), \mathbf{A}(\mathbf{r,t}),$ and $\hat\Delta(\mathbf{r}, \mathbf{r}')$, then $\ket{u(r)}\exp[-i\chi(\mathbf{r})]$ and $\ket{v(r)}\exp[i\chi(\mathbf{r})]$ are solutions with potentials $V(\mathbf{r}), e\mathbf{A}(\mathbf{r,t})/\hbar + \nabla \chi(\mathbf{r})$ and $\hat\Delta(\mathbf{r}, \mathbf{r}')\exp[-i(\chi(\mathbf{r})+\chi(\mathbf{r}'))]$. The proof is by direct substitution.

We focus on uniform vector potentials, and treat vector potential $\mathbf{A}(t)$ as a \textit{parameter} in Eqs.(\ref{eq:realspacebdg},\ref{eq:realspacebdg2}). Since the translational symmetry is preserved, we employ the Bloch theorem to find the solutions of Eqs.(\ref{eq:realspacebdg},\ref{eq:realspacebdg2}) with the finite supercurrent in the following form:
\begin{equation}
    \ket{\mathcal{U}(\mathbf{r})} = \ket{\mathcal{U}_\mathbf{k,q}(\mathbf{A})} e^{i(\mathbf{k}+\mathbf{q})\cdot \mathbf{r}},\qquad \ket{\mathcal{V}(\mathbf{r})} = \ket{\mathcal{V}_\mathbf{k,q}(\mathbf{A})} e^{i(\mathbf{k}-\mathbf{q})\cdot \mathbf{r}}.
    \label{eq:uvbloch}
\end{equation}
We introduce the $\mathbf{k}$-Hamiltonian as  $\hat{H}^{(0)}_{\mathbf{k}}[\mathbf{A}(t)]=e^{-i\mathbf{k}\cdot \mathbf{r}}\hat{H}_0(\mathbf{r},-i\boldsymbol{\nabla}+e\mathbf{A}(t)/\hbar)e^{i\mathbf{k}\cdot \mathbf{r}}$ and by substitution of Eq.~(\ref{eq:uvbloch}) into Eqs.(\ref{eq:realspacebdg},\ref{eq:realspacebdg2}) we find the familiar $\mathbf{k}$-space BdG equations:
\begin{align}
    \hat{H}^{(0)}_{\mathbf{k}+\mathbf{q}}[\mathbf{A}(t)]&\ket{\mathcal{U}_\mathbf{k,q}(\mathbf{A})} +\hat{\Delta}_{\mathbf{k},\mathbf{q}}\ket{\mathcal{V}_\mathbf{k,q}(\mathbf{A})} = E\ket{\mathcal{U}_\mathbf{k,q}(\mathbf{A})},\label{kBdG1}\\-\{\hat{H}^{(0)}_{-\mathbf{k}+\mathbf{q}}[\mathbf{A}(t)]\}^*&\ket{\mathcal{V}_\mathbf{k,q}(\mathbf{A})} +\hat{\Delta}^\dagger_{\mathbf{k},\mathbf{q}}\ket{\mathcal{U}_\mathbf{k,q}(\mathbf{A})} = E\ket{\mathcal{V}_\mathbf{k,q}(\mathbf{A})}\label{kBdG2}
\end{align}
where $\hat{\Delta}_{\mathbf{k},\mathbf{q}}=\int d\mathbf{r}'\hat{\Delta}(\mathbf{r}, \mathbf{r}')e^{i\mathbf{k}\cdot (\mathbf{r}'-\mathbf{r})}$ and note, $[H^{(0)}_{\mathbf{k}}]^* = [H^{(0)}_{\mathbf{k}}]^T$. For local pairing $\hat{\Delta}(\mathbf{r}, \mathbf{r}')= \hat{\Delta}\delta(\mathbf{r}- \mathbf{r}')$, we write $\hat{\Delta}_{\mathbf{k},\mathbf{q}} = \hat{\Delta}$. We emphasise that $\frac{e}{\hbar}\mathbf{A}$ enters the \textit{periodic parts} of Eq.(\ref{eq:uvbloch}) on an equal footing as $\mathbf{q}$, which can be seen after direct substitution of Eq.(\ref{eq:uvbloch}) into Eqs.(\ref{eq:realspacebdg},\ref{eq:realspacebdg2}). Thus, the derivatives with respect to $\frac{e}{\hbar}\mathbf{A}$ of the lattice periodic quantities is equivalent to derivatives with respect to $\mathbf{q}$; for instance, $\partial_A^\alpha \ket{\mathcal{U}_\mathbf{k,q}(\mathbf{A})} \equiv \frac{e}{\hbar}\partial_q^\alpha \ket{\mathcal{U}_\mathbf{k,q}(\mathbf{A})}$. 

For simplicity we consider a system with no spin orbit coupling $H_{\mathbf{k},\uparrow}^{(0)} = H_{\mathbf{k},\downarrow}^{(0)}$. Due to the structure of the BdG equations as in Eqs.(\ref{kBdG1}-\ref{kBdG2}), and the fact that both $\ket{m,\mathbf{k+q}}$ and $\ket{m,\mathbf{-k+q}}^*$ form a full basis at every $\mathbf{k}$, it is convenient to expand the $\ket{\mathcal{U}_\mathbf{k,q}}$ in $\ket{m,\mathbf{k+q}}$ basis while $\ket{\mathcal{V}_\mathbf{k,q}}$ in $\ket{m,\mathbf{-k+q}}^*$. The solution of the BdG equations can be thus simplified by using the \textit{isolated band approximation} as:
\begin{equation}\label{genIBO}
    \ket{\mathcal{U}_\mathbf{k,q}(\mathbf{A})} = \bar{\mathcal{U}}_{\mathbf{k,q}}(\mathbf{A})\ket{n,\mathbf{k+q}+\frac{e}{\hbar}\mathbf{A}},\qquad \ket{\mathcal{V}_\mathbf{k,q}(\mathbf{A})} = \bar{\mathcal{V}}_{\mathbf{k,q}}(\mathbf{A})\ket{n,\mathbf{-k+q}+\frac{e}{\hbar}\mathbf{A}}^*,
\end{equation}
leading to the following, simplified (reduced), BdG equations:
\begin{equation}\label{reducedBdG}
    \left(\begin{array}{cc}
        \epsilon_{\mathbf{k+q}+\frac{e}{\hbar}\mathbf{A}} & \Delta(\mathbf{k,q}+\frac{e}{\hbar}\mathbf{A}) \\
        \Delta^*(\mathbf{k,q}+\frac{e}{\hbar}\mathbf{A}) & -\epsilon_{\mathbf{-k+q}+\frac{e}{\hbar}\mathbf{A}}
    \end{array}\right)\left(\begin{array}{c}
         \bar{\mathcal{U}}_{\mathbf{k},\mathbf{q}}(\mathbf{A})  \\
         \bar{\mathcal{V}}_{\mathbf{k},\mathbf{q}}(\mathbf{A})
    \end{array}\right)=\mathcal{E}_{\mathbf{k}}(\mathbf{q}+\frac{e}{\hbar}\mathbf{A})\left(\begin{array}{c}
         \bar{\mathcal{U}}_{\mathbf{k},\mathbf{q}}(\mathbf{A})  \\
         \bar{\mathcal{V}}_{\mathbf{k},\mathbf{q}}(\mathbf{A})
    \end{array}\right),
\end{equation}
where $\Delta_{\mathbf{k,q+\frac{e}{\hbar}\mathbf{A}}} = \bra{n,\mathbf{k+q}+e\mathbf{A}/\hbar}\hat{\Delta}(\ket{n,\mathbf{-k+q}+e\mathbf{A}/\hbar}^*) = |\Delta(\mathbf{k},\mathbf{q}+e\mathbf{A}/\hbar)|e^{i{\rm arg}[\Delta_{\mathbf{k,q+\frac{e}{\hbar}\mathbf{A}}}]}$. Importantly, the $\mathbf{k,q,A}$ dependence of the superconducting gap $\Delta_{\mathbf{k},\mathbf{q}}(\mathbf{A})$ appears due to the projection of the multiband Hamiltonian onto the shifted Bloch states at a Fermi surface $\ket{n_{\rm FS},\pm\mathbf{k+q}+e\mathbf{A}/\hbar}$. The multiband order parameter matrix $\hat\Delta$ is $\mathbf{q,A}$ independent (for local pairing also $\mathbf{k}$ independent). Only the \textit{kinetic} part of the multiband BdG Hamiltonian is coupled to the vector potential $\mathbf{A}(t)$. 

We find the lower Bogoliubov state of Eq.(\ref{reducedBdG}) as $ [v_{\mathbf{k},\mathbf{q}}(\mathbf{A})  , u_{\mathbf{k},\mathbf{q}}(\mathbf{A})]^{\rm T}$ 
where $u_{\mathbf{k},\mathbf{q}}(\mathbf{A}) =|u_{\mathbf{k},\mathbf{q}}(\mathbf{A}) |, v_{\mathbf{k},\mathbf{q}}(\mathbf{A}) =-|v_{\mathbf{k},\mathbf{q}}(\mathbf{A}) |e^{i{\rm arg}[\Delta_{\mathbf{k,q+\frac{e}{\hbar}\mathbf{A}}}]}$ with $2|v_{\mathbf{k,q}}(\mathbf{A})|^2 = 1- \tilde{\epsilon}_{\mathbf{k},\mathbf{q}}(\mathbf{A})/[(\tilde{\epsilon}_{\mathbf{k},\mathbf{q}}(\mathbf{A}))^2+4|\Delta_{\mathbf{k},\mathbf{q}}(\mathbf{A})|^2]^{1/2}$ , $2|u_{\mathbf{k,q}}(\mathbf{A})|^2 = 1+ \tilde{\epsilon}_{\mathbf{k},\mathbf{q}}(\mathbf{A})/[(\tilde{\epsilon}_{\mathbf{k},\mathbf{q}}(\mathbf{A}))^2+4|\Delta_{\mathbf{k},\mathbf{q}}(\mathbf{A})|^2]^{1/2}$,  $\tilde{\epsilon}_{\mathbf{k},\mathbf{q}} = \epsilon_{\mathbf{k+q+e\mathbf{A}/\hbar}}+\epsilon_{\mathbf{-k+q+e\mathbf{A}/\hbar}}$ and the energy of the lower Bogoliubov band is $2\mathcal{E}_{\mathbf{k},-}(\mathbf{q}+\frac{e}{\hbar}\mathbf{A}) = \epsilon_{\mathbf{k+q+e\mathbf{A}/\hbar}}-\epsilon_{\mathbf{-k+q+e\mathbf{A}/\hbar}}- [(\tilde{\epsilon}_{\mathbf{k},\mathbf{q}}(\mathbf{A}))^2+4|\Delta_{\mathbf{k},\mathbf{q}}(\mathbf{A})|^2]^{1/2}$. Note, $v_{\mathbf{k},\mathbf{q}}(\mathbf{A})  , u_{\mathbf{k},\mathbf{q}}$ are the standard coherence factors found in the superconducting state displayed in Eq.(\ref{SCGS}) of the main text. For the TRS systems, the isolated band approximation in Eq.(\ref{genIBO}) reduces to expression used by Ref.\cite{Liang_2017}.  

$\mathcal{D}^{\alpha\beta \gamma}$ principally requires the breaking of inversion symmetry and appears even in time reversal symmetric systems; it persists when time-reversal symmetry is broken. However, if time reversal and inversion symmetry are simultaneously broken while their composite inversion and time-reversal symmetry (PT) is preserved, $\mathcal{D}^{\alpha\beta \gamma}$ identically vanishes~\cite{PhysRevLett.127.277202}. This means that both P and PT need to be broken for $\mathcal{D}^{\alpha\beta \gamma}$ to manifest. Notice, that such symmetry constraints apply to the entire interacting phase that includes both the lattice as well as the order parameter. In addition to these general symmetry requirements, in 2D systems rotations also constrain $\mathcal{D}^{\alpha\beta \gamma}$; finite $\mathcal{D}^{\alpha\beta \gamma}$ in 2D requires breaking all rotational symmetries. As a result, mirror symmetry is the largest point group symmetry compatible with $\mathcal{D}^{\alpha\beta \gamma}$ in a 2D material~\cite{SodemannFu}. For a mirror axis along $x$, we find $\mathcal{D}^{x xx}=\mathcal{D}^{x yy} =\mathcal{D}^{y xx}=\mathcal{D}^{y yy} = \mathcal{D}^{yyx}= \mathcal{D}^{yxy} = 0$ and $\mathcal{D}^{x xy} = -\mathcal{D}^{x yx}$.

\subsection*{3. Rhombohedral trilayer graphene with strain.}
For the results shown in Fig.~\ref{sBCDFIG2} of the main text, we focus on a strained rhombohedral trilayer graphene~\cite{Jung_2013}:
\begin{equation}
\label{eq:sixbandrtg}
    H_{\rm ABC} =\left(\begin{array}{ccccccc}
         u_1+\delta_1& \frac{1}{2}\gamma_2 & v_0\pi^\dagger & v_4\pi^\dagger & v_3\pi -\gamma_N& v_6\pi\\
         \frac{1}{2}\gamma_2& u_3+\delta_1 & v_6\pi^\dagger & v_3\pi^\dagger & v_4\pi&v_0\pi \\
         v_0\pi& v_6\pi &u_1 &\gamma_1 &v_4\pi^\dagger &v_5\pi^\dagger \\
         v_4\pi& v_3\pi &\gamma_1 &u_2 &v_0\pi^\dagger &v_4\pi^\dagger \\
         v_3\pi^\dagger - \gamma^*_N& v_4\pi^\dagger & v_4\pi &v_0\pi &u_2 &\gamma_1 \\
         v_6\pi^\dagger& v_0\pi^\dagger & v_5\pi & v_4\pi&\gamma_1 & u_3 
    \end{array}\right)\quad \text{in}\quad \left[\begin{array}{c}
          A_1\\
          B_3\\
          B_1\\
          A_2\\
          B_2\\
          A_3\\
    \end{array}\right]
\end{equation}
with $\pi_\xi = \xi k_x + ik_y$ where $\xi = \pm1$ is the valley index and the strain modifies ($A_1,B_2$) elements according to Ref.\cite{Jung_2013}. Parameters will be listed shortly. Similarly to Ref.\cite{Zhang_2010} we set $v_5=v_6=0$ and assume quantities $v_0\pi$,$u_1,u_2,v_3,v_4,\gamma_N$ are small. 

For computational simplicity, in our work we will focus on an effective low-energy two band Hamiltonian obtained in a L\"owdin decomposition of the six-band hamiltonian in Eq.~(\ref{eq:sixbandrtg}). Gathering the leading contributions we find the effective two-band Hamiltonian in $(\ket{A_1},\ket{B_3})$ basis \cite{Zhang_2010,Jung_2013} as:
\begin{multline}\label{ABC}
    \hat{H}_{\rm RTG}^{(0)}(\mathbf{k},\xi)=\frac{v_0^3}{\gamma_1^2}\left(\begin{array}{cc}
        0 &(\pi_\xi^\dagger)^3  \\
        (\pi_\xi)^3 & 0
    \end{array}\right)+\frac{v_0\gamma_N }{\gamma_1} \left(\begin{array}{cc}
        0 &\pi_\xi^\dagger  \\
         \pi_\xi & 0
    \end{array}\right)+\delta_2\left(1-3\frac{k^2v^2_0}{\gamma^2_1}\right)\tau_0+\delta\left(1-\frac{k^2 v_0^2}{\gamma_1^2}\right)\tau_z\\+\left(\delta_1-\frac{2v_4v_0k^2}{\gamma_1}\right)\tau_0 +\left(\frac{1}{2}\gamma_2 -\frac{2k^2v_0v_3}{\gamma_1} \right)\tau_x, 
\end{multline}
with crystal momentum $k=|\mathbf{k}|$, $\gamma_N(\mathbf{k})=\gamma_{N0}e^{- v_0 k/\gamma_1}$, with $\gamma_1/v_0 = 0.0573\text{\AA}^{-1}$ and $v_i = {\sqrt{3} a \gamma_i}/{2},~a = 2.46 \text{\AA},\delta = (u_1-u_3)/2,3\delta_2=(u_1+u_3)/2-u_2$. In the main text, we used the notation $g(\mathbf{k}) = 1-{k^2 v_0^2}/{\gamma_1^2}$ and $\mathbf{d}(\mathbf{k})\cdot\boldsymbol{\tau} \equiv d_0(\mathbf{k})\tau_0+d_x(\mathbf{k})\tau_x+d_y(\mathbf{k})\tau_y+d_z(\mathbf{k})\tau_z$. Note, in the main text we omitted the valley index $\xi$ for brevity; they are shown here. The derived full Hamiltonian preserves time-reversal symmetry, while $\delta$ can be controlled by external gate potential breaks inversion symmetry and the strain $\gamma_N$ breaks $C_{3z}$ \cite{Jung_2013}. Parameters are adopted from Refs.\cite{Zhang_2010,Jung_2013} in eV as:

\begin{table}[h]
    \centering
    \begin{tabular}{c|c|c|c|c|c|c|c|c}
    \hline
        $\gamma_0$ &$\gamma_1$  &$\gamma_2$  &$\gamma_3$  &$\gamma_4$   &$\delta$ &$\delta_1$  &$\delta_2$&$\gamma_{N0}$  \\\hline3.1&0.38&-0.015&0.29&0.141&0.03&-0.0105&-0.0023&0.02\\\hline
    \end{tabular}
    \caption{Table of parameters for rhombohedral trilayer graphene~\cite{Zhang_2010,Jung_2013}.}
\end{table}

\subsection*{4. Bogoliubov de Gennes equations, self consistency, and model parameters.} We adopt the standard mean-field Bogoliubov de Gennes description of a superconductor with a local attractive density-density interaction potential $\hat{V}_{\rm el-el}^{ss'}(\mathbf{r}-\mathbf{r}')=-U \delta(\mathbf{r}-\mathbf{r}')\delta_{s,\bar{s}'}$, where $s,\bar{s}$ are the opposite spins, due to the Pauli blocking. The supercurrent is introduced by modelling the order parameter with a real space dependence as $\Delta(\mathbf{r})=\Delta e^{2i\mathbf{q}\cdot\mathbf{r}}$. We employ local in space constant pairing~Ref.\cite{LevitovSC}, where the pairing occurs between the electrons of opposite momenta ($\mathbf{k}$ of $K_+$ valley with $-\mathbf{k}$ of $K_-$ valley). Note our treatment is agnostic to the mechanism for attraction; nevertheless we note that in RTG, attraction between electrons has been proposed to originate from soft critical fluctuations near isospin-polarized states in rhombohedral trilayer graphene~\cite{LevitovSC}.

We account for the effect of electron-electron interactions in the self-consistent Hartree-Fock fashion by self-consistently solving $\Delta_{\mathbf{q},ab} = -U\sum_{\mathbf{k}}\langle c_{-\mathbf{k}+\mathbf{q},a}c_{\mathbf{k}+\mathbf{q},b}\rangle$ with $a,b$ to be the sub-lattice degrees of freedom. Such treatment is equivalent to a Free energy minimization. Writing the Greens function of the BdG system as $\mathcal{G}^{\rm BdG}_{\mathbf{k},\mathbf{q},\omega_n}=[i\hbar\omega_n - \hat{H}^{\rm BdG}_{\mathbf{k,q},0}]^{-1}$ \cite{Altland_Simons_2010}, with Fermionic Matsubara frequency $\hbar\omega_n = \pi\beta^{-1}(2n+1)$, the Free energy reads as:
\begin{equation}\label{scdelta}
    F_{\mathbf{q}} = \frac{\Delta_0^2}{2U} -\frac{1}{2\beta}\sum_{k,n} {\rm Tr ln}[\mathcal{G}^{\rm BdG,-1}_{\mathbf{k},\mathbf{q},\omega_n}]\quad \Rightarrow \quad \frac{\partial F(\Delta_A,\Delta_B)}{\partial \Delta_i} = 0\quad\Rightarrow\quad\frac{\Delta_i}{U} =-\frac{1}{2}\sum_{k} {\rm Tr}\{f(\hat{H}^{\rm BdG}_{\mathbf{k,q},0} )\partial_{\Delta_i} \hat{H}^{\rm BdG}_{\mathbf{k,q},0} \}, 
\end{equation}
where we used $ \partial_x {\rm ln} [A^{-1}(x)] = A(x)\partial_x A^{-1}(x)$ and $\sum_{n}\frac{1}{i\hbar\omega_n - E_a}=f(E_a)$. Note, for the effective model described in the previous section, summation over $\mathbf{k}$ in the self-consistent scheme above includes summation over both valleys leading to a single global $\hat\Delta$. Due to the TRS, summation over valleys leads to extra factor of 2 in a self-consistent procedure of Eq.(\ref{scdelta}).  We write the mean-field Bogoliubov-de-Gennes (BdG) Hamiltonian for each valley as:
\begin{equation}\label{ABCBdG}
    \hat{H}^{\rm BdG}_{\mathbf{k,q},0}(\xi) = \left( \begin{array}{cc}
        \hat{H}_{\rm RTG}^{(0)}(\mathbf{p+q},\xi)-\mu\tau_0 & \hat{\Delta} \\
        \hat{\Delta}^\dagger & -\hat{H}_{\rm RTG}^{*(0)}(\mathbf{-p+q},-\xi) +\mu\tau_0
    \end{array}\right).
\end{equation}
Where we adopted the spin blind Hamiltonian. In our simulations we self-consistently determined the order parameter $\Delta(\mathbf{q})$ for small $\mathbf{q}$. We observed both $\Delta_A$ and $\Delta_B$ to be even functions of $\mathbf{q}$ and plateaus at small $\mathbf{q}$, which is the limit of interest in this work. 

In order to study the the superconducting BCD, we examined superconducting RTG. We adopted $\mu\approx 17$ meV (point of high DoS) $\delta = 0.03$ meV and a cut-off frequency of $\hbar\omega_D = 40$ meV for the self-consistent treatment.  For illustration we used $U = 60$meV yielding a modest $T_c\approx 1$ K. We note, parenthetically, $T_c$ in graphene stacks can be controlled e.g., in hybrid graphene stack layered with a transition metal dichalcogenides~\cite{SCenhancement}. Self consistently determined order parameter at $T =0$ K is $\Delta_A \approx 1.5\cdot10^{-1}$ meV, $\Delta_B \approx 10^{-2}$meV. We find the gap opening in the spectrum of the BdG Hamiltonian of order $\Delta_{\rm SC} \approx 2\cdot10^{-1}$ meV. As a result, in Fig.\ref{sBCDFIG2}a we used $\hat{\Delta} \approx [0.15,0;0,0.01]$ meV, $\delta = 0.03$ meV, $\mu\approx 17$ meV. 

To illustrate the proximity effect from a centrosymmetric superconducting on noncentrosymmetric RTG in Fig.\ref{sBCDFIG2}b we chose $\hat{\Delta} \approx [0.15,0;0,0.15]$ meV, $\delta = 0.03$ meV, $\mu\approx 17$ meV. Similarly, to illustrate the proximity effect of a noncentrosymmetric superconductor in centrosymmetric target layer (in this case RTG) in Fig.\ref{sBCDFIG2}c we adopted $\hat{\Delta} \approx [0.4,0;0,0.04]$ meV, $\delta = 0$ meV, $\mu=-1$ meV. All other parameters were kept the same. 

\subsection*{5. Quantum Liouville evolution and multiband linear and nonlinear response of superconductors.}
To study multiband superconductor response, we analyze the superconducting state via a Bogoliubov de Gennes (BdG) Hamiltonian $\hat{H}= \frac{1}{2}\sum_{\mathbf{k}}\Psi^\dagger_{\mathbf{k,q}} \hat{H}^{\rm BdG}_{\mathbf{k,q}}[\mathbf{A}(t)]\Psi_{\mathbf{k,q}}$ in the Bloch-Nambu basis $\Psi_{\mathbf{k,q}}=(\vec{c}_{\mathbf{k}+\mathbf{q}},\vec{c}^{~\dagger}_{-\mathbf{k}+\mathbf{q}})^{\mathrm{T}}$ (note that this vectorial arrangement of $\vec{c}$ runs over all degrees of freedom e.g. spin, sub-lattice)
\begin{equation}
    \hat{H}^{\rm BdG}_{\mathbf{k,q}}[\mathbf{A}(t)] \hspace{-1mm}=\hspace{-1mm}{\left(\begin{array}{cc}
         \hat{H}_{\mathbf{k+q}}^{(0)}[\mathbf{A}(t)]&\hat \Delta  \\
         \hat{\Delta}^\dagger&-\{\hat{H}^{(0)}_{\mathbf{-k+q}}[\mathbf{A}(t)]\}^T
    \end{array}\right)},
    \label{eq:BdGHSI}
\end{equation}
where the Bloch Hamiltonian of the parent material is $H_{\mathbf{k}}^{(0)} =  H_0(\mathbf{k}) - \mu\mathbb{I}$ and the pairing operator is $\hat{\Delta}$. Note that coupling to electromagnetic $\mathbf{A}(t)$ proceeds in the standard fashion through the kinetic part of the BdG Hamiltonian \cite{DeGennes}. Here $\mu$ is the chemical potential and $H_{\mathbf{k}}^{(0)} \ket{n, \mathbf{k}} = \epsilon_{\mathbf{k}} \ket{n, \mathbf{k}}$. In contrast, $\ket{\Psi_a}$ eigenstates of the BdG Hamiltonian in Eq.~(\ref{eq:BdGHSI}) describe Bogoliubov quasiparticles that comprise the superconducting state. 

In this section we systematically derive the photo-response of a multiband superconductor by employing the Hermitian Liouville equation for the Bogoliubov quasiparticles $i\hbar \partial_t\hat{\rho}(t)= [\hat{H}^{\rm BdG}_{\mathbf{k,q}}[\mathbf{A}(t)],\hat{\rho}(t)],$ where $\hat\rho(t)$ is the time-dependent Bogoliubov quasiparticle density matrix. The most general expansion of the BdG Hamiltonian in powers of electric field can be performed as:
\begin{equation}\label{eq:hBdGexp}
    \hat{H}^{\rm BdG}_{\mathbf{k,q},t} = H^{\rm BdG}_{\mathbf{k},\mathbf{q}}+\frac{e}{\hbar}A^\alpha_t \partial_\mathbf{q}^\alpha \hat{H}^{\rm BdG}_{\mathbf{k},\mathbf{q}}+\frac{1}{2!}\frac{e^2}{\hbar^2}A^\alpha_t A^\beta_t \partial_\mathbf{q}^\alpha \partial_\mathbf{q}^\beta \hat{H}^{\rm BdG}_{\mathbf{k},\mathbf{q}}+\frac{1}{3!}\frac{e^3}{\hbar^3}A^\alpha_t A^\beta_t A^\gamma_t\partial_\mathbf{q}^\alpha \partial_\mathbf{q}^\beta \partial_\mathbf{q}^\gamma \hat{H}^{\rm BdG}_{\mathbf{k},\mathbf{q}}+\mathcal{O}(E^4).
\end{equation}
where $\hat{H}^{\rm BdG}_{\mathbf{k,q}}[0]\equiv \hat{H}^{\rm BdG}_{\mathbf{k,q}}$ with $\hat{H}^{\rm BdG}_{\mathbf{k,q}} \ket{\Psi_a(\mathbf{k,q})} = \mathcal{E}_{a,{\mathbf{q}}}(\mathbf{k})\ket{\Psi_a(\mathbf{k,q})}$. Further we omit explicit indication of dependence on $\mathbf{k,q}$ for compactness.

Next we aim to clarify the matrix elements in the Bogoliubov basis for each of the terms in 
Eq.~(\ref{eq:hBdGexp}). We start from the considering the following quantity:
\begin{equation}
     \bra{\Psi_a}(\partial_\mathbf{q}^\alpha O)\ket{\Psi_b}=\partial_\mathbf{q}^\alpha(\bra{\Psi_a}O\ket{\psi_b})-\bra{\partial_\mathbf{q}^\alpha\Psi_a} O\ket{\Psi_b}-\bra{\Psi_a} O\ket{\partial_\mathbf{q}^\alpha\Psi_b}= \partial^{\alpha}_\mathbf{q}O_{ab}-i[\mathcal{A}^\alpha,O]_{ab},
\end{equation}
where $\mathcal{A}^\alpha_{ab} = \bra{\Psi_a}i\partial^\alpha_{\mathbf{q}}\Psi_b\rangle$, we used $\partial^\alpha_\mathbf{q} (\bra{\Psi_a}\Psi_b\rangle)=0$ and the resolution of identity in the Bogoliubov states so that $[A,B]_{ab}\equiv\sum_c A_{ac}B_{cb}-B_{ac}A_{cb}$. Applying this procedure multiple times, we find the following identities:
\begin{align}
    &h^\alpha_{ab}=\bra{\Psi_a}(\partial_\mathbf{q}^\alpha H^{\rm BdG}_{\mathbf{k},\mathbf{q}})\ket{\Psi_b}=\partial^\alpha_\mathbf{q}\mathcal{E}_{\mathbf{q},a} (\mathbf{k})\delta_{ab} + i[\mathcal{E}_{\mathbf{q},a} (\mathbf{k})-\mathcal{E}_{\mathbf{q},b} (\mathbf{k})]\mathcal{A}^\alpha_{ab}  ,\label{eq:va}\\
    &h^{\alpha\gamma}_{ab}=\bra{\Psi_a}(\partial_\mathbf{q}^\alpha\partial_\mathbf{q}^\gamma H^{\rm BdG}_{\mathbf{k},\mathbf{q}})\ket{\Psi_b} = \partial^{\alpha}_\mathbf{q}h^{\gamma}_{ab}- i[\mathcal{A}^\alpha,h^{\gamma}]_{ab} ,\label{eq:vab}\\
    &h^{\alpha\beta\gamma}_{ab}=\bra{\Psi_a}(\partial_\mathbf{q}^\alpha \partial_\mathbf{q}^\beta \partial_\mathbf{q}^\gamma H^{\rm BdG}_{\mathbf{k},\mathbf{q}})\ket{\Psi_b}=\partial^\beta_\mathbf{q} h^{\alpha\gamma}_{ab} - i[\mathcal{A}^\beta,h^{\alpha\gamma}]_{ab}. \label{eq:vabc}
\end{align} 

By adopting the adiabatic turning on regularization for the electric field $A^\alpha_t=e^{\eta t}\int_\omega A^\alpha_\omega e^{-i\omega t}$ and using the BdG Hamiltonian expansion in Eq.(\ref{eq:hBdGexp}), we follow the standard perturbation theory and iteratively solve the Liouville equation to find the first order density matrix elements in the Bogoliubov basis as:
\begin{equation}
    [\hat\rho_1(t)]_{ab} = \int_\omega e^{-i\omega t+\eta t/\hbar}\frac{e}{\hbar}A^\alpha_{\omega}\frac{h^\alpha_{ab}f_{ba}}{\hbar\omega -\mathcal{E}_{ab}+ i\eta} ,
\end{equation}
where $f_{ab}\equiv f[\mathcal{E}_{\mathbf{q},a} (\mathbf{k})]-f[\mathcal{E}_{\mathbf{q},b} (\mathbf{k})], \mathcal{E}_{ab}\equiv\mathcal{E}_{\mathbf{q},a} (\mathbf{k})-\mathcal{E}_{\mathbf{q},b} (\mathbf{k})$ and the second order as:
\begin{equation}
    [\hat\rho_{2}(t)]_{ab} =\iint_{\omega_1\omega_2} \frac{e^{-i\omega_1 t -i\omega_2 t+ 2\eta t/\hbar}\frac{e}{\hbar}A^\alpha_{\omega_1}\frac{e}{\hbar}A^\beta_{\omega_2}}{\hbar\omega_1+\hbar\omega_2-\mathcal{E}_{ab}+ 2i\eta}\left[\frac{1}{2}h^{\alpha\beta}_{ab}f_{ba}+ \sum_c\frac{h^\beta_{ac}h^\alpha_{cb}f_{bc}}{\hbar\omega_1 -\mathcal{E}_{cb}+ i\eta}- \sum_c\frac{h^\alpha_{ac}h^\beta_{cb}f_{ca}}{\hbar\omega_1 -\mathcal{E}_{ac}+ i\eta} \right].
\end{equation}

Next, the electric current is evaluated as the trace of the electric current operator $\hat{j}^\alpha=\partial H/ \partial A^\alpha$ with the density matrix: $j^\gamma = \int_{\mathbf{k}}{\rm Tr}[\hat{j}^\gamma \hat{\rho}]$ where $\int_{\mathbf{k}}\equiv\int{d\mathbf{k}}/{(2\pi)^2}$. Keeping terms up to the second order in $A$ produces:
\begin{equation}
    j^\alpha(t) = \frac{e}{\hbar}\int_{\mathbf{k}}{\rm Tr}\bigg[\left(\hat h^\alpha+\frac{e^2}{\hbar^2}A^\beta \hat h^{\alpha\beta}+\frac{1}{2}\frac{e^3}{\hbar^3}A^\beta A^\gamma \hat h^{\alpha\beta\gamma}\right)(\hat \rho_0+\hat \rho_1+\hat \rho_2)\bigg]+\mathcal{O}(E^3).
\end{equation}
Thus we identify the \textit{linear response} as:
\begin{equation}
    j_{(1)}^\alpha(t) =    \frac{e^2}{\hbar^2}\int_\omega e^{-i\omega t+\eta t/\hbar}A^\beta_{\omega}\int_{\mathbf{k}} \sum_{ab}\frac{h^\beta_{ab}h^\alpha_{ba}f_{ba}}{\hbar\omega -\mathcal{E}_{ab}+ i\eta} +\frac{e^2}{\hbar^2}A^\beta_t\int_{\mathbf{k}}\sum_a f_a  h_{aa}^{\alpha\beta}.
\end{equation}
\begin{figure*}
\centering
\includegraphics[width=0.98\textwidth]{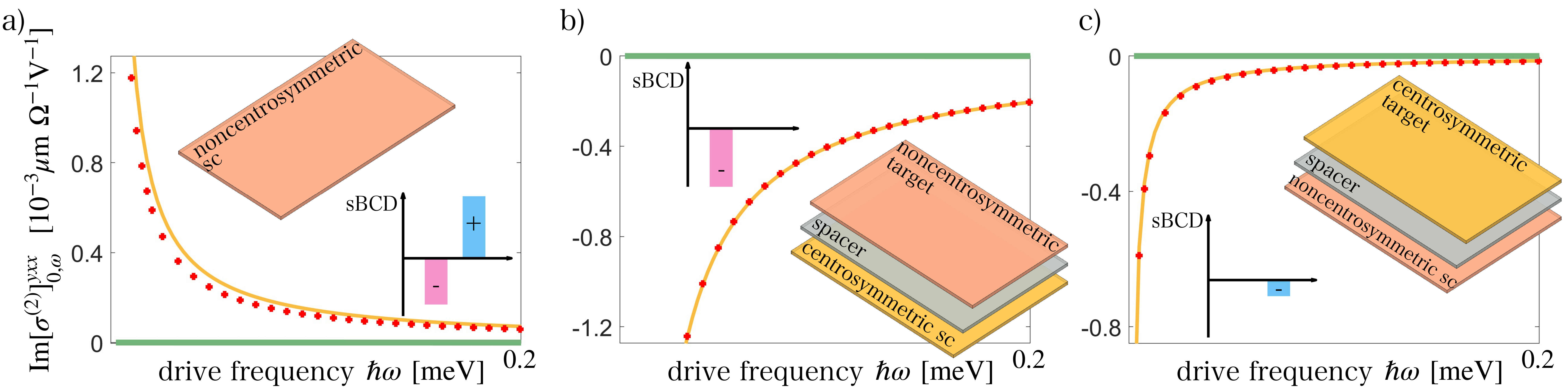}
\caption{{\it Superconducting Berry curvature dipole second order nonlinear responses.} Second order nonlinear conductivity as a function of frequency $\omega$; $\mathcal{D}_{\rm sBCD}$ second order nonlinearity (orange line) and full multiband response (red dots, see Section 5 of SM) in three setups: {\bf a.} non-centrosymmetric superconductor with a noncentrosymmetric orbital structure for the superconductor pairing $\hat\Delta = [\Delta_{A},0;0,\Delta_{B}]$ {\bf b.} non-centrosymmetric target layer with a centrosymmetric proximity induced pairing structure $\hat\Delta_0^{\rm proximity} = [\Delta_{0},0;0,\Delta_{0}]$ {\bf c.} centrosymmetric target layer with a noncentrosymmetric proximity effect: pairing possesses a non-trivial orbital structure $\hat\Delta^{\rm proximity} = [\Delta^{\rm proximity}_{A},0;0,\Delta^{\rm proximity}_{B}]$. To provide a comparison for the giant sBCD nonlinear response, we compared it against second-order nonlinearities in the parent metal with $\tau = 1$ ps (green).}
\label{sBCDFIG2SI}
\end{figure*}

Using  Eqs.(\ref{eq:vab}) and decomposing partial fractions we obtain:
\begin{equation}\label{SIconductivity}
    [\sigma^{(1)}]^{\alpha\beta}_\omega=- i\frac{e^2}{\hbar}\int_{\mathbf{k}}\left[\frac{1}{\hbar\omega+i\eta}\sum_a f_a\partial^\beta h^\alpha_{aa} + \sum_{ab}\frac{h^\beta_{ab}h^\alpha_{ba}f_{ab}}{\hbar\omega -\mathcal{E}_{ab}+ i\eta}\frac{1}{\mathcal{E}_{ba}}\right],
\end{equation}
while the \textit{non-linear response} is found to have the following four components:
\begin{align}
    &{j}^{\gamma,1}_{(2)}(t)=\frac{e^3}{\hbar^3}\int_{\mathbf{k}}\int_{\omega_1,\omega}e^{-i\omega t+ 2\eta t/\hbar} A^\alpha_{\omega_1}A^\beta_{\omega-\omega_1}\sum_a\frac12 f_ah^{\alpha\beta\gamma}_{aa},\label{eq:j1}\\
    &{j}^{\gamma,2}_{(2)}(t)=\frac{e^3}{\hbar^3}\int_{\mathbf{k}}\int_{\omega_1,\omega} e^{-i\omega t+ 2\eta t/\hbar}A^\alpha_{\omega_1}A^\beta_{\omega-\omega_1}\sum_{abc}\frac{h^\alpha_{c b}h^\beta_{ac}h^\gamma_{ba}}{\hbar\omega-\mathcal{E}_{ab}+2i\eta}\Bigg[\frac{ f_{bc}}{\hbar\omega_1-\mathcal{E}_{cb}+i\eta}+\frac{ f_{ac}}{\hbar\omega-\hbar\omega_1-\mathcal{E}_{ac}+i\eta}\Bigg]\label{70},\\
    &{j}^{\gamma,3}_{(2)}(t)=\frac{e^3}{\hbar^3}\int_{\mathbf{k}}\int_{\omega_1,\omega} e^{-i\omega t+ 2\eta t/\hbar}A^\alpha_{\omega_1}A^\beta_{\omega-\omega_1}\sum_{ab} \frac{f_{ba}h^{\alpha}_{ab}h^{\beta\gamma}_{ba}}{\hbar\omega_1-\mathcal{E}_{ab}+i\eta}\label{71},\\
    &{j}^{\gamma,4}_{(2)}(t)=\frac{e^3}{\hbar^3}\int_{\mathbf{k}}\int_{\omega_1,\omega} e^{-i\omega t+ 2\eta t/\hbar}A^\alpha_{\omega_1}A^\beta_{\omega-\omega_1}\sum_{ab} \frac{\frac12f_{ba}h^{\alpha\beta}_{ab}h^{\gamma}_{ba}}{\hbar\omega-\mathcal{E}_{ab}+2i\eta},\label{eq:j4}
\end{align}
which is consistent with Refs.\cite{Yanase1,Yanase2}. The explicit expressions for $\hat h^{\alpha},\hat h^{\alpha\beta},\hat h^{\alpha\beta\gamma}$ can be found in Eqs.~(\ref{eq:va}) to (\ref{eq:vabc}) above. This multiband formulation based off the BdG equations are shown as the red dots in the main text.  In obtaining of red dots for the Fig. 2 of the main text we focused on Eq.(\ref{SIconductivity}) and expanded to the lowest order in $\mathbf{q}$ in the limit $\hbar\omega\ll \Delta$. 

We note that Refs.~\cite{Yanase1,Yanase2} massaged Eq.~(\ref{70}) and (\ref{71}) to produce a low-frequency sBCD nonlinear response  $[\sigma^{(2)}]_{\omega, \omega_1}$ discussed in the main text. Note that care, however, should be applied when manipulating such BdG multiband nonlinear response expressions. For example, if the isolated band approximation is directly applied to Eq.~(\ref{70}) and (\ref{71}), the low-frequency response vanishes even when the superconducting BCD discussed in the main text is finite. Instead, a summation across all bands should be performed before applying the isolated band approximation. 

Nevertheless, it is remarkable that the sBCD nonlinear response based on the BCS superconducting state used in the main text, corresponds so well to a BdG multiband approach ~\cite{Yanase1,Yanase2}. Indeed, in Fig.\ref{sBCDFIG2SI}, we plot the nonlinear response from  $\ket{\rm SC_\mathbf{q}}$ state based results (yellow lines) as discussed in the main text. We find it closely aligns with the multiband formalism (red dots). 

Importantly, such sBCD induced second-order nonlinearity is {\it giant} as it diverges for small $\omega$ far exceeding the values expected in the metallic state (green line). This makes sBCD induced second-order nonlinearities a choice probe of quantum geometric quantities superconductors and renders noncentrosymmetric superconductors an attractive platform to exploit quantum geometric nonlinearities for superconducting circuits.  

\subsection*{6. Line cut of order parameter sBCD.}
\begin{figure}[t]
    \centering
    \includegraphics[width=0.3\textwidth]{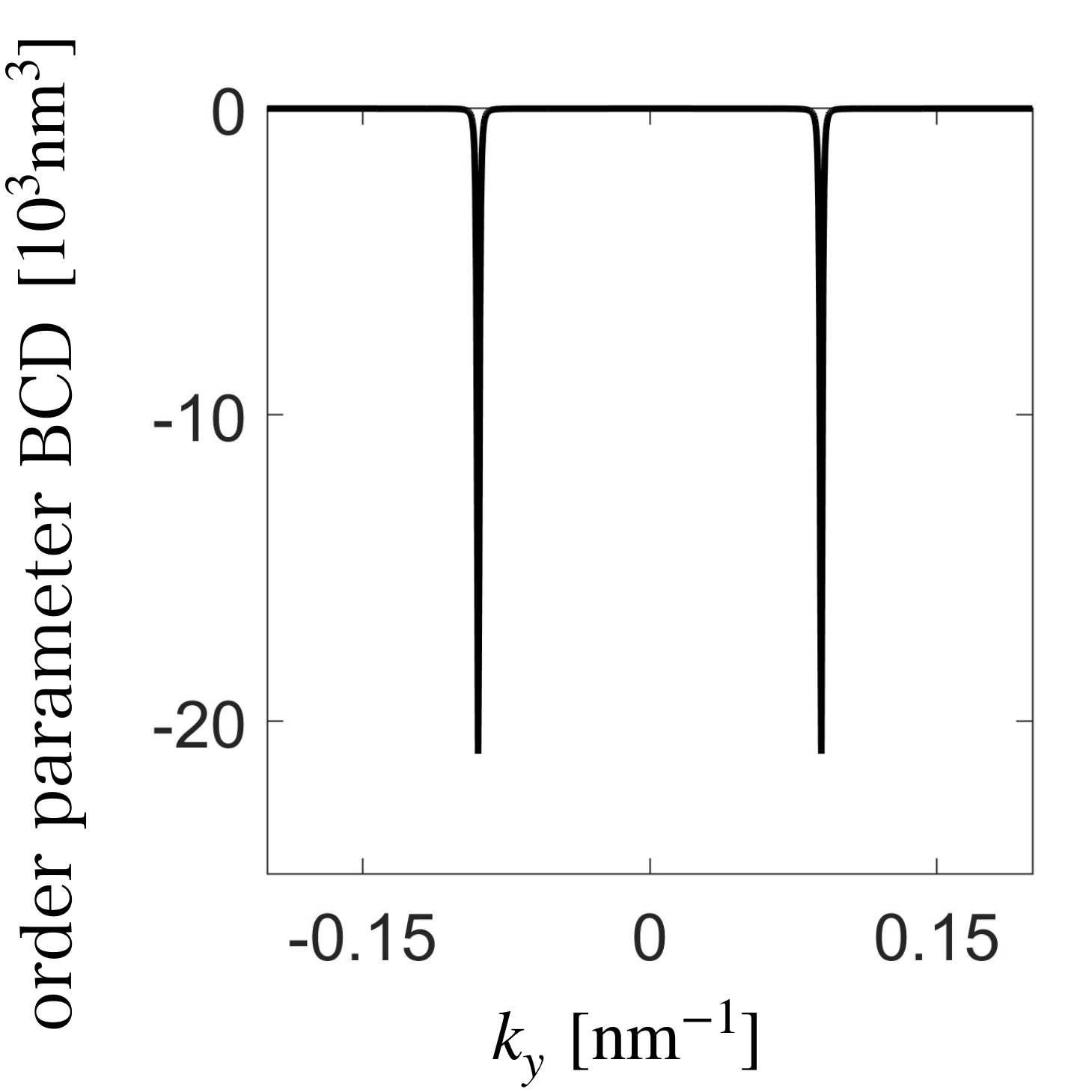}
    \caption{$k$-space slice of the order parameter Berry curvature dipole distribution for superconducting RTG as on FIG.\ref{sBCDFIG1}{\bf d} at $k_x=0.2$ [nm$^{-1}$]. Notice the distribution is highly peaked close to the BdG band edges.}  
    \label{sBCDFIG1_slice}
\end{figure}

 In the main text, we displayed order parameter sBCD using a saturated colorbar in order to make the contours of where it occurred more apparent for the reader. Here we show a line cut that display its full scale height in Fig.~\ref{sBCDFIG1_slice}. 

\end{document}